\documentclass[12pt]{article}
\usepackage[pdftex]{graphicx,color}
\usepackage{amsmath,amssymb}
\usepackage{amsfonts}
\usepackage{graphicx,color}
\usepackage{cite}
\usepackage{bm,bbm}
\usepackage{comment}

\newcommand{\nn}{\nonumber}

\newcommand{\be}         {  \begin{equation}  }
\newcommand{\ee}           {  \end{equation}  }
\newcommand{\eqb}{\begin{eqnarray}}
\newcommand{\eqe}{\end{eqnarray}}
\def\ba  { \begin{align} }
\def\ea  { \end{align} }

\newcommand{\eqn}[1]{(\ref{#1})}

\def\ep{\epsilon}

\def\bra{\langle}
\def\ket{\rangle}
\def\cdots {\cdot\cdot\cdot}
\def\pint#1 {- \!\!\!\!\!\!\!\! \,\int_{#1}}
\def\ni       {\noindent}

\def\comma      { \, , }
\def\period     { \, . }
\def\del        {  \partial  }
\def\delbar     {  \bar{\partial}  }
\def\half       {\textstyle \frac{1}{2}}
\def\abs#1      {  \, \vert #1 \vert \,   }
\def\binom#1#2 { \vecii{ {}_{#1} }{\raisebox{.5ex}{$ {}^{#2} $}} }

\def\zbar   {\bar{z}}
\def\cbar   {\bar{c}}

\newcommand{\bbR}{{\mathbb R}}
\newcommand{\bbZ}{{\mathbb Z}}

\DeclareMathOperator{\re}{Re}
\DeclareMathOperator{\diag}{diag}

\def\wvec#1#2    {  \left(\begin{array}{c}#1\\#2\end{array}\right)  }
\def\vecii#1#2     {{ #1 \choose #2 }  }
\def\veciii#1#2#3  {\left(\begin{array}{c}#1\\#2\\#3\end{array}\right)}
\def\matrixii#1#2#3#4   {\begin{pmatrix} #1 & #2 \\ #3 &  #4\end{pmatrix}  }
\def\smallmatrixii#1#2#3#4 { \left( \begin{smallmatrix} #1& #2 \\ #3 & #4 \end{smallmatrix} \right) }
\def\matrixiii#1#2#3#4#5#6#7#8#9 {\left(\begin{array}{ccc}#1&#2&#3\\
                                #4&#5&#6\\#7&#8&#9\end{array}\right)}
\renewcommand{\thesection}
{\arabic{section} \hspace{-.5em}}
\renewcommand{\thesubsection}
{\arabic{section}.\arabic{subsection}   \hspace{-.5em}}
\renewcommand{\thesubsubsection}
{\arabic{section}.\arabic{subsection}.\arabic{subsubsection} \hspace {-.5em} }

\makeatletter
\renewcommand\section{
\@startsection{section}{3}{\z@}%
{-3.25ex\@plus -1ex \@minus -.2ex}%
{1.5ex \@plus .2ex}%
{\normalfont\large\bfseries\mathversion{bold}}}
\renewcommand\subsection{
\@startsection{subsection}{3}{\z@}%
{-3ex\@plus -1ex \@minus -.2ex}%
{1.5ex \@plus .2ex}%
{\normalfont\normalsize\bfseries\mathversion{bold}}}
\renewcommand\subsubsection{
\@startsection{subsubsection}{3}{\z@}%
{-3.25ex\@plus -1ex \@minus -.2ex}%
{1.5ex \@plus .2ex}%
{\normalfont\normalsize\itshape}}
\makeatother

\makeatletter
\renewcommand{\appendix}{
\renewcommand{\thesection}{\Alph{section}  \hspace{-.5em}}
\renewcommand{\thesubsection}
{\Alph{section}.\arabic{subsection} \hspace{-.5em}}
\renewcommand{\thesubsubsection}
{\Alph{section}.\arabic{subsection}.\arabic{subsubsection} \hspace  {-.5em}}
\setcounter{section}{0}
}
\makeatother
\def\titleandfile#1#2#3   {  \begin{center}{\large\bf #1}\end{center}
                            \par\begin{flushright} #2 \\ #3  \end{flushright} 
                          }
\renewcommand{\thefootnote}{\fnsymbol{footnote}}

\def\calC  {{\cal C}}

\def\calJ  {{\cal J}}
\def\calM  {{\cal M}}

\def\qbar  { \bar{q} }
\def\zetabar  { \bar{\zeta} }
\def\ubar  { \bar{u} }
\def\kappabar  { \bar{\kappa} }

\def\Rb { R_{5} }

\def\papertitlepage{\baselineskip 3.5ex \thispagestyle{empty}}

\renewcommand{\thefootnote}{\fnsymbol{footnote}}

\makeatletter \@addtoreset{equation}{section} \makeatother
\renewcommand{\theequation}{\arabic{section}.\arabic{equation}}

\renewcommand{\thefootnote}{\arabic{footnote}}
\begin{document}
\papertitlepage
\setcounter{page}{0}


\vspace*{3ex}
\baselineskip=4ex
\begin{center}
{\Large\bf\mathversion{bold}
Interactions of strings on a T-fold
}
\end{center}

\vskip 0.5 ex 
\baselineskip=3.6ex
\begin{center}
  Yuji Satoh\footnote[1]{\tt ysatoh@u-fukui.ac.jp} 
  and
  Yuji Sugawara\footnote[2]{\tt ysugawa@se.ritsumei.ac.jp}
 
\vskip 4ex
$^\ast${\it Department of Applied Physics, University of Fukui}\\
 {\it Bunkyo 3-9-1, Fukui 910-8507,  Japan}\\

\vskip 3ex
$^\dagger${\it Department of Physical Sciences, College of Science and Engineering}\\
 {\it Ritsumeikan University, Shiga 525-8577, Japan}
 
\end{center}

\baselineskip=3.6ex
\vskip 8ex
\begin{center} {\bf Abstract} \end{center}
\vskip 1ex
We consider the interactions of strings on T-folds 
from the world-sheet point of view which are exact in $\alpha'$.
As a concrete example, we take a model 
where the internal torus at the 
so(8) enhancement point is twisted by T-duality (T-folded),
and compute the scattering amplitudes of a class of massless strings.
The four-point amplitudes involving both twisted and untwisted strings
are obtained in a closed form in terms of the hypergeometric function.
By their factorization, 
the three-point coupling of the twisted and untwisted strings
is found to be suppressed by the chiral momenta 
along the internal torus, 
and quantized in integer powers of 1/4.
The asymptotic forms of the four-point amplitudes in high-energy limits
are also obtained.
Our results rely only on general properties 
of the asymmetric orbifold by the T-duality twist and 
of the Lie algebra lattice from the symmetry enhancement,
and thus  may be extended qualitatively to more general T-folds.

\vskip 3ex
\vspace*{\fill}
\noindent
March, 2022
%
%

\setlength{\parskip}{1ex}
\renewcommand{\thefootnote}{\arabic{footnote}}
%
\setcounter{footnote}{0}
\setcounter{section}{0}
\pagestyle{plain}

\baselineskip=3.6ex
\newpage

\section{Introduction}
\label{sec:introduction}

Strings can propagate in backgrounds which are not 
geometric in the conventional sense. 
Although each appropriate conformal field theory (CFT) 
on the world-sheet
gives a string vacuum,
its target-space interpretation is not always obvious. 
It is also possible to think of the backgrounds where 
the transition functions for the target space 
involve the symmetries intrinsic to string theory, namely, the string dualities.
In the case of T-duality, the corresponding ``(non-)geometry''  is named 
the T-fold \cite{Hellerman:2002ax,Dabholkar:2002sy,Flournoy:2004vn,Hull:2004in}.
These non-geometric backgrounds are relevant in understanding
 the string vacua. Furthermore, 
 those involving the string dualities would provide
  clues in understanding the dualities and the new formulations
 of string theory, such as Double Field Theory (DFT) \cite{Hull:2009mi}, 
 in which the dualities are manifest. 
   
Among non-geometric backgrounds in string theory, we focus on 
the T-fold in this paper. At low-energy,
strings on T-folds can be analyzed by supergravity, or 
under certain consistent truncation including the winding modes,
by DFT.  Moreover,
in order to study them in the regime where their quantum effects 
are further incorporated,
their formulation based on the world-sheet CFT would be necessary.
Since the T-folding is realized on the world-sheet by gauging 
a twist involving T-duality which acts asymmetrically on the left- 
and the right-mover,
T-folds are described by a particular class of asymmetric 
orbifold CFTs.%
\footnote{For the world-sheet doubled formalism, 
see \cite{Hull:2004in,Hull:2006va,Berman:2013eva}.
For a review on non-geometric backgrounds in string theory in general, see e.g.
\cite{Plauschinn:2018wbo}.
}

In such a formulation, it has been known that one encounters 
a seeming puzzle 
\cite{Aoki:2004sm,Flournoy:2005xe,Hellerman:2006tx,Kawai:2007qd}: 
To be concrete, let us 
consider the T-duality transformation which is realized 
by a chiral reflection of the coordinate fields 
e.g. in the right-mover,  and hence acts as a $\bbZ_2$
transformation. The modular transformation 
of the world-sheet partition functions, however, generally forms 
a $\bbZ_{n>2 }$ representation due to the non-trivial phases in the twisted 
sectors which remain as a consequence of the left-right asymmetry. 
A resolution to this puzzle is to note that the $\bbZ_2$ action 
in the target space may be lifted to more general actions 
on the world-sheet 
\cite{Aoki:2004sm,Satoh:2015nlc,Satoh:2016izo,Harvey:2017rko}.
In other words, T-duality may have a different representation there,
similarly to the spin representation for the orthogonal group.

Once the puzzle is cleared, it is conceptually straightforward to 
advance the studies of strings on T-folds by 
the world-sheet theories, though it is still non-trivial 
in practice 
since the construction and the analysis of the corresponding 
asymmetric orbifold theories are rather involved. 
In this respect, a class of the asymmetric orbifold
CFTs describing T-folds, or their modular invariant partition functions,
has been constructed 
in a systematic manner by making use of Lie algebra lattices 
\cite{Satoh:2015nlc,Satoh:2016izo}. 
Motivated by the cosmological constant problem as in 
\cite{Satoh:2015nlc}, related asymmetric orbifold models 
have been constructed \cite{Sugawara:2016lpa,Aoyama:2020aaw,Aoyama:2021kqa}.
An application of these CFTs/models  to that problem 
has  been discussed \cite{Satoh:2021nfu}.

The purpose of this paper is to make a step toward the
studies of strings on T-folds in depth which are exact in $\alpha'$ 
beyond the regime  of supergravity and DFT. In particular, 
we take a supersymmetric model
constructed in \cite{Satoh:2015nlc} as a concrete
example, where the internal torus at the 
so(8) enhancement point is twisted by T-duality (T-folded), 
and study the interactions of the strings.
Our prime interest is to figure out
what happens when the strings twisted by T-duality,
which might be rather exotic from the conventional 
point of view, interact with the untwisted strings.
As simple examples to this end, we consider 
a class of ten-dimensional massless strings, 
which are excited by the internal momenta above 
the four-dimensional massless strings in the untwisted sector
or those in the ground states in the twisted sector.
We first  construct their vertex operators.
The T-duality twist, or the chiral reflection in this case, is implemented 
by the twist fields of the type 
 in the Ashkin-Teller model 
 \cite{Bershadsky:1986mt,Dixon:1986qv,Zamolodchikov:1987ae,
 Ginsparg:1988ui,Hashimoto:1996he,Frohlich:1999ss,Mukhopadhyay:2001ey,
 Mattiello:2018kue}
 for the bosonic coordinate fields and by
 the spin fields for the world-sheet fermions.
 
 We then compute their three- and four-point amplitudes,
 focusing on the NS-NS sector. The four-point amplitudes 
 involving two twisted and two untwisted strings
 are obtained in a closed form in terms of the hypergeometric function.
  By their factorization, we find
 that the three-point amplitudes among the twisted and untwisted
 strings are suppressed by the momenta flowing along the T-folded
 internal torus. Its mechanism is essentially the same as
 that for general symmetric orbifolds discussed in \cite{Hamidi:1986vh}.
 In our case of an asymmetric orbifold based on a Lie algebra lattice, 
 only the momenta in the right-mover contribute 
 to the suppression. 
 In addition, the three-point coupling is quantized 
in integer powers of $N^{-2}$ with $N=2$, where $N$ comes from 
  $\bbZ_{N=2}$ of the T-duality twist (in the untwisted sector) 
 and the exponent  2 from the length of 
  the roots of so(8).
 This result of the suppression 
 includes stringy effects through the winding modes which are
non-perturbative from the sigma-model point of view.
 
 We also consider high-energy limits of the amplitudes, which 
 correspond to $\alpha' \to \infty$ 
 opposite to the particle limit  $\alpha' \to 0$.
 In a hard-scattering limit where (the absolute values of)
 all the Mandelstam variables become large,
the angle dependence for the momenta along the twisted internal torus
takes the same form as that for the external momenta
in the usual hard-scatting. The sign of its exponent
given by the magnitude of the incident momenta is, however, 
opposite.
We find that both of the two saddle points of the integrand of the amplitude
are dominant, and each gives the correct result in the hard-scattering 
limit up to a momentum dependent phase.
 
Although we work with a concrete model of strings on T-folds,
 our analyses rely only on general properties of the asymmetric orbifold
 by the T-duality twist and of the Lie algebra lattice associated with 
 the twisted internal torus. Thus,  our results may be extended
  qualitatively
 to more general T-folds.
 
The rest of this paper is organized as follows.
In section 2, we review a model constructed in \cite{Satoh:2015nlc},  
which is used 
in this paper, and summarize its properties. This section sets up our
notation. In section 3, we classify the spectrum of the model,
and construct the vertex operators 
of the strings mentioned above.
In section 4, we compute their three- and four-point amplitudes 
in the case of the vanishing 
right-moving momenta along the T-duality twisted torus.
This section serves as a preparation for the next section, and 
provides a check of the results there as well.
In section 5, we compute the amplitudes in the general case 
with the non-vanishing right-moving momenta along the twisted torus.
We analyze their properties including the suppression of the coupling 
and the high-energy behaviors. We conclude with a summary and discussion 
in section 6. In the appendix,  we summarize our conventions, 
some formulas and details of our computations, 
which are used in the main text.

\section{A model of strings on T-folds}
\label{sec:ourmodel}

\subsection{T-fold from so(8) torus}
\label{subsec:Tfoldso8}
\vspace{-1ex}

As a concrete example, we consider a model of strings on T-folds which is
constructed in \cite{Satoh:2015nlc}. This model is obtained 
by gauging a twist of
$\calM = \bbR^{1,3} \times S^1 
\times \bbR_{\rm base} \times T^4_{\rm fiber}$
which involves T-duality, 
where 
$T^4_{\rm fiber} $ is the torus at the so(8) 
enhancement point of the moduli. 
We denote the coordinate fields by 
$X^M$ $(M=0,1, ..., 9)$. 
For later use, we introduce the indices for each part  of $\calM$; for example, 
$\tilde\mu = M= 0, ..., 3$ for $\bbR^{1,3}$,
$p = M= 4,5$ for $S^1$ and $\bbR_{\rm base}$ respectively,
and $a=M=6,...,9$ for $T^4_{\rm fiber}$.
The part other than $T^4_{\rm fiber}$ is also denoted, e.g. 
by $\mu = M = 0, ..., 5$.
In the untwisted sector, the twist $\sigma$ for our T-fold is implemented by
the T-duality twist or the chiral reflection
 on $T^4_{\rm fiber}$,
\be
\label{crefX}
 (X_L^a, X_R^a)~ \longmapsto ~ (X_L^a, -X_R^a)
 \qquad (a=6,7,8,9) \comma
\ee
together with the translation in $\bbR_{\rm base}$,
\be
\label{basetrans}
X^5  \longmapsto   X^5 + 2 \pi \Rb \comma 
\ee
where we have split $X^a$ into the  left- and the right-mover
as $X^a(z,\zbar) = X_L^a(z) + X_R^a(\zbar) $.
Thus, a string wrapped around the base circle along $\bbR_{\rm base}$ 
modded out by \eqn{basetrans} undergoes monodromy due to \eqn{crefX}
in the fiber torus part $T^4_{\rm fiber}$, 
which characterizes the T-fold.  It turns out, as in \eqn{Ztw} below, that 
half-integer modes, in addition to integer ones, appear 
in the spectrum of the momentum along the base circle.%
\footnote{
One may start with a base circle with radius $2 R_5$
as in \cite{Flournoy:2004vn,Hellerman:2006tx,Kawai:2007qd}
instead of $\bbR_{\rm base}$, 
and then consider the T-duality twist accompanied with the half-shift 
$X^5\, \longmapsto \, X^5 + 2\pi R_5$.
This is equivalent to the present formulation and leads 
to the same partition function as in \eqn{Ztw}.
}

These $X^a_{L/R}$ are expanded as
\be
  X^a_{L} = 
  e_j^a  x^j_{L}  - i \frac{\alpha'}{2} p^a_L \ln z + \cdots
  \comma \quad
   X^a_{R} = 
   e_j^a  x^j_{R}  - i \frac{\alpha'}{2} p^a_R \ln \zbar + \cdots
   \comma
\ee
with 
\be
\label{pLR}
  p_{L}^a = \sqrt{2 \over \alpha'} \, e^{\ast j}_a \big[n_j +(G-B)_{jk} w^k \big]
  \comma \quad
  p_{R}^a = \sqrt{2 \over \alpha'} \,  e^{\ast j}_a \big[n_j -(G+B)_{jk} w^k \big]
 \comma
\ee
and $n_j, w^j \in \bbZ$.
Here,   $e^a_j$ and  $e^{*j}_a$ are the vielbein and its inverse, which satisfy
\begin{align}
\label{vielbein}
 &  e^{a}_j e^{a}_k = 2G_{jk} \comma \quad
 e^{\ast j}_a e^{\ast k}_a = \frac{1}{2} G^{jk} \comma \quad
  e^{a}_j e^{\ast j}_b  = \delta^a_b  \period 
\end{align}
In the present case, the 
constant space-time metric $G_{ij}$ and the 
anti-symmetric tensor $B_{ij}$ are given by 
the Cartan matrix $C_{ij}$ of so(8) as 
$ G_{ij} = \frac{1}{2} C_{ij} $ for any $i,j$ and $B_{ij} = \frac{1}{2} C_{ij}$ for $i>j$.
$G^{ij}$ is the inverse of $G_{ij}$. 
Thus,
$e^a_j$ and $e^{*j}_a$  form  the root and the weight lattice of so(8), respectively.
This also implies that  the momenta $\sqrt{\alpha'/2} \, p^a_{L/R}$ take 
their value
on the so(8) lattice. 
The left- and right-momenta belong to the same conjugacy class,
since  their difference is given by the root lattice,  
$\sqrt{\alpha'/2} (p^a_{L} - p^a_{R}) = e^{a}_j w^j$.
We follow the notation in \cite{Giveon:1994fu,Satoh:2016izo}
for the Lie algebra lattice. 
The periodicity of 
$\hat{X}^j := e^{* j}_a X^a \sim \hat{X}^j + \sqrt{\alpha'/2} \times 2\pi m^j $
with $m^j \in \bbZ$
is translated into $X^a \sim  X^a +\sqrt{\alpha'/ 2} \times 2 \pi e^a_j \, m^j $.
The chiral bosons $X^M_{L/R}$ have the standard 
operator product expansions (OPEs),
$X_L^M(z)X_L^N(0) \sim - (\alpha'/2) \eta^{MN}\ln z $ and 
$X_R^M(\zbar)X_R^N(0) \sim - (\alpha'/2) \eta^{MN}\ln \zbar $
with $\eta^{MN} = \diag(-1, +1, ..., +1)$.

For the world-sheet superconformal symmetry to be preserved, 
the T-duality twist acts also as the chiral reflection
on the  world-sheet 
fermions in the untwisted Neveu-Schwarz (NS) sector,
\be
 \label{crefpsi}
   (\psi_L^a, \psi_R^a)~ \longmapsto ~ (\psi_L^a, -\psi_R^a)
 \qquad (a=6,7,8,9) \period
\ee
The fermions in the left- and the  right-mover, 
$\psi^M_{L}(z)$, $\psi^M_{R}(\zbar)$,  have
the OPEs,  $\psi_L^M(z) \psi_L^N(0) \sim \eta^{MN}z^{-1} $ and 
$\psi_R^M(\zbar) \psi_R^N(0) \sim \eta^{MN}\zbar^{-1} $ .
The action (\ref{crefpsi}) 
determines the T-duality twist in this sector.
That in 
the untwisted Ramond sector is, however, not unique
and there are essentially two cases, as understood 
e.g. through bosonization. When one adopts a
bosonization of the right-moving transverse fermions $\psi_R^M$,%
\footnote{
We have omitted the cocycles to ensure the anti-commutativity
among different fermions.
}
\begin{eqnarray}
\label{psiRbasis}
&& \psi^2_R \pm i \psi^3_R =: \sqrt{2} e^{\pm iH^1_{R}},
\qquad 
\psi^4_R \pm i \psi^5_R =: \sqrt{2} e^{\pm iH^2_{R}}, \nonumber \\
&& \psi^6_R \pm i \psi^7_R =: \sqrt{2} e^{\pm iH^3_{R}},
\qquad 
\psi^8_R \pm i \psi^9_R =: \sqrt{2} e^{\pm iH^4_{R}} \comma
\end{eqnarray}
by the free bosons $H_R^k(\zbar)$ with 
$H_R^k(\zbar) H_R^l(0) \sim - \delta^{kl} \ln \zbar$, 
the action (\ref{crefpsi}) is translated into
\begin{equation}
\label{crefH}
 (H^1_{R}, H^2_{R}, H^3_{R}, H^4_{R} ) ~ \longmapsto ~ 
(H^1_{R}, H^2_{R}, H^3_{R} + \pi , H^4_{R} - \pi) \period
\end{equation}
In terms of these $H^k_{R}$, the spin fields are given by
\be
\label{defSR}
S_R(\ep) = \check{S}_R \hat{S}_R
 \comma 
\quad \check{S}_R:=e^{\frac{i}{2} \sum_{k=1}^2 \ep_k H^k_{R}}
\comma \quad
\hat{S}_R := e^{\frac{i}{2} \sum_{k=3}^4 \ep_k H^k_{R}}
\comma
 \ee
with $\ep_k = \pm 1$.
We denote the spin fields also by using the spinor indices e.g. as 
$S_R^\alpha$ instead of $S_R(\ep)$.
Then, the T-duality twist squares to the identity, 
and thus the twist becomes $\bbZ_2$ also in this sector.

Alternatively, 
one can  bosonize $\psi_R^M$ so that two of 
$\psi_R^a$ $(a=6, ..., 9)$ for $T^4_{\rm fiber}$
are mixed with those for the other part $\psi_R^M$ $(M=2, ..., 5)$. 
In this case, 
(\ref{crefpsi}) results in a $\bbZ_4$ action in
the untwisted Ramond sector.
In the following, we consider the first case in \eqn{psiRbasis}, \eqn{crefH}.

\subsection{Partition function and spectrum}

The actions (\ref{crefX}), (\ref{basetrans}), (\ref{crefpsi}), (\ref{crefH}) 
determine the twist $\sigma$ for the T-fold
in the untwisted sector. Together with
the Gliozzi-Scherk-Olive (GSO) projected untwisted partition function, 
they also determine
the partition function twisted once by $\sigma$. 
Those in the twisted sectors 
are obtained from them by modular  transformations. 
Consequently, the partition function for the transverse part 
 becomes 
\begin{align}
\label{TfoldZ}
Z(\tau, \bar{\tau}) = & 
Z^{\rm tr}_{\bbR^{1,3}\times S^1}(\tau,\bar{\tau}) 
 \, \calJ (\tau)\, Z_{\rm tw}(\tau, \bar{\tau}) \comma \nn \\
 & Z_{\rm tw}(\tau, \bar{\tau}) = 
  \sum_{w,m \in \bbZ}\,  Z_{(w,m)}^{\rm base}(\tau,\bar{\tau}) \,
F^{T^4}_{(w,m)}(\tau, \bar{\tau}) \,
 \overline{f_{(w,m)}(\tau)} 
 \period
\end{align} 
Here, $\tau$ is the modulus of the world-sheet torus.
$Z^{\rm tr}_{\bbR^{1,3}\times S^1}$, $ Z_{(w,m)}^{\rm base}$, 
$F^{T^4}_{(w,m)}$ are the partition functions 
from $X^\mu$
 $(\mu=2,3,4)$ for the transverse part of $\bbR^{1,3}\times S^1$,
 from $X^5$ for $\bbR_{\rm base}$, and from 
 $X^a$ $(a=6,...,9)$ for $T^4_{\rm fiber}$, respectively.
 $ \calJ$ comes from the left-moving fermions
 $\psi_L^M$ $(M=2, ..., 9)$, whereas
 $\overline{f_{(w,m)}}$ from the right-moving fermions
 $\psi_R^M$ $(M=2, ..., 9)$.
 The explicit form of $Z^{\rm tr}_{\bbR^{1,3}\times S^1}$ 
 is the standard one, which we omit.
 Those for the other components are listed in appendix \ref{app:partfn}.
 
We note that the action of the T-duality
twist on the world-sheet is generally
uplifted 
\cite{Aoki:2004sm,Satoh:2015nlc,Satoh:2016izo,Harvey:2017rko}.
In the present  case,
it becomes 
 $\bbZ_4$ on the internal $T^4_{\rm fiber}$ part as 
in (\ref{FabT4}) and (\ref{fab}) due to the phases depending on
$w$ and $m$,
though it remains $\bbZ_2$ in the untwisted sector, i.e. 
the sector with $w =0$.   
The action 
in the twisted sectors, 
i.e. the sectors with $w \neq 0$, is found from 
the expression of the partition function.

This model preserves  24 space-time supercharges
(3/4 supersymmetry), 16 of which come from the left-mover
and 8 of which from the right-mover.  Its spectrum
is read off by the Poisson resummation with respect to 
the temporal winding $m$, which converts $m$ to the momentum 
$n$
along $X^5 \in S^1$. After some algebra, one finds
\begin{align}
\label{Ztw}
&Z_{\rm tw}(\tau, \bar{\tau})  \\
&  =
\frac{1}{|\eta(\tau)|^2}
  \sum_{w,n \in \bbZ} 
 q^{\frac{\alpha'}{4} \bigl( \frac{n}{2\Rb} + \frac{\Rb}{\alpha'}w \bigr)^2}
 \qbar^{\frac{\alpha'}{4} \bigl( \frac{n}{2\Rb} - \frac{\Rb}{\alpha'} w\bigr)^2}
  \sum_{m \in \bbZ_2} (-1)^{nm}
 F^{T^4}_{(w,m)}(\tau, \bar{\tau}) \overline{f_{(w,m)} (\tau)}
  \comma \nn
\end{align}
where 
$q=e^{2\pi i \tau}$ and $\eta(\tau)$ is the Dedekind $\eta$ function.
The spectrum is tachyon-free and maintains the unitarity.
Once the phases in $F^{T^4}_{(w,m)} $ and $ \overline{f_{(w,m)}} $
are combined and cancelled with each other, 
the action of the T-duality twist
on the product $ F^{T^4}_{(w,m)} \, \overline{ f_{(w,m)}} $ 
becomes $\bbZ_2$ in our case. 
In general, the T-duality twist on the internal part, however,  
remains to be $\bbZ_k$ with $k>2$.
In the target space, T-folds generally have non-geometric fluxes ($Q$-fluxes)
 \cite{Dabholkar:2005ve}.  Those in this model may be read off 
 by following a general procedure given in  \cite{Condeescu:2013yma} 
which realizes non-geometric fluxes in terms of world-sheet variables.

\section{Vertex operators} 
\label{sec.vertexop}

\subsection{Classification of spectrum by conjugacy classes}
The spectrum in the above partition function
can be classified by the conjugacy classes of 
the so($2k$) lattice. 
We denote by o, v, s, c the conjugacy class including
the vacuum, the vector, the spinor and the conjugate-spinor representation,
respectively. 
The characters of o+v, o$-$v, s+c, s$-$c
are represented by  
the theta functions $\theta_3^k$,  $\theta_4^k$, 
$\theta_2^k$,
$\vartheta_1^k := (-i\theta_1)^k$ divided by $\eta^k$.

In the left-mover, $\calJ(\tau)$ contains v and s of so(8).
In the right-mover,  the contents of $f_{(w,m)}$ read
\begin{align}
f_{(0,0)}  :  & \ \  {\rm (ov + vo) - (ss+cc) } \comma  \qquad
f_{(0,1)}  :   \ \   {\rm (-ov + vo) - (ss-cc) } \comma \nn \\
f_{(1,0)}  :  & \ \  {\rm (os + vc) - (sv+co) } \comma  \qquad 
f_{(1,1)}  :   \ \   {\rm (os - vc) - (-sv+co) } \period 
\end{align}
Here, we have decomposed the conjugacy classes of so(8)
to those of so(4) for  $\psi_R^\mu$ $(\mu=2, ..., 5)$
and those of so(4) for $\psi_R^a$ $(a=6, ..., 9)$,
and denoted them from the left to the right.
For example, 
ov means o for $\psi_R^\mu$  and 
v for $\psi_R^a$.

As a consequence of the so(8) symmetry in the untwisted sector,
 $F^{T^4}_{(w,m)} $ are also given by the theta functions
as in \eqn{FabT4}, and accordingly
decomposed by the conjugacy classes as
\begin{align}
\label{Fab}
F^{T^4}_{(0,0)}  :  & \ \   {\rm o\,\overline{(oo + vv)} + v \,\overline{(ov + vo)}
+ s\, \overline{(ss+cc)} + c\, \overline{(sc+cs)}} 
\comma \nn \\
F^{T^4}_{(0,1)}  :   & \ \  {\rm o \,\overline{(oo - vv)} + v \, \overline{(-ov + vo)}
+ s\, \overline{(ss-cc)} + c\, \overline{(-sc+cs)} } 
\comma \nn\\
F^{T^4}_{(1,0)}  :  & \ \   {\rm o\, \overline{(oc + vs)} +v\, \overline{(os + vc)} 
+ s\, \overline{(sv+co)} + c\, \overline{(so+cv)}} 
\comma \\ 
F^{T^4}_{(1,1)}  :   & \ \   
{\rm o\, \overline{(oc - vs)} + v\, \overline{(-os + vc) }
+ s\, \overline{(sv-co)} + c\, \overline{(-so+cv)}} 
\period  \nn
\end{align}
Here, 
the left-moving  part is represented by the conjugacy classes  of so(8),
whereas 
the right-moving part is represented by  those of 
so(4)$\,\oplus\,$so(4)
 and denoted with the overline $\overline{(\ \ )}$.
While the action of the twist on the untwisted sector directly follows 
from its definition, the one on the twisted sector is determined
uniquely by the modular invariance of the total partition function and 
the modular covariance of its building blocks.
The actions of the twist in the twisted and untwisted sectors are 
generally different as above.

\subsection{Massless strings in the untwisted sector}
 \label{sec:utmassless}

In $F^{T^4}_{(0,m)}$ for the untwisted sector, the conjugacy 
classes are specified by the momenta $p_L$ and $p_R$
along $T^4_{\rm fiber}$.  
Under the twist $\sigma$ for the T-fold, the states 
may have minus signs 
from the excitations of the oscillators
$\tilde\alpha_{-k}^a$ $(k \in \bbZ_{>0})$ of $X_R^a$,
the odd combination
of their oscillator vacua $\vert p_R \rangle - \vert -p_R \rangle$,
and/or the momentum part for $\bbR_{\rm base}$ 
with $n \in 2\bbZ+1$ in \eqn{Ztw}.
The states invariant under the twist are obtained 
in such a way that these signs are cancelled with each other.

In the following, we consider 
the ten-dimensional massless strings which belong to
the untwisted sector (of the internal $T_{\rm fiber}^4$ part) 
with $w, n\in 2 \bbZ$ in (\ref{Ztw}),
as simple examples to probe the interactions on the T-fold.
With the left-moving fermions included,
the relevant part of the partition function reads
$  \big(F^{T^4}_{(0,0)}\overline{f_{(0,0)}}
 + 
  F^{T^4}_{(0,1)} \overline{f_{(0,1)}} \big) \calJ$.
We also concentrate on the NS-NS sector.
Massless strings involving the Ramond sectors are related
 by supersymmetry.
The conjugacy class including the NS-NS massless states 
is then labeled as 
\be
\label{ut0}
    \bigm[ {\rm o\, \overline{(oo)}}  \bigm]_X \times  
    \bigm[{\rm v\, \overline{(vo)} } \bigm]_\psi
\period
\ee
The part inside $[\quad]_X$ is from $F^{T_4}_{(w,m)}$.
For the fermion part $[\quad]_\psi$,  
we have denoted the conjugacy classes 
in the order of $[\psi_L^M  (\psi_R^\mu, \psi_R^a) ] $ 
$( M= 2, ...., 9; \, \mu = 2, ...,5; \, a = 6, ..., 9)$.
In the states of the massless strings, the excited non-zero modes 
are $(\psi_L)^M_{-1/2}$ and $(\psi_R)^\mu_{-1/2}$ only.
Before taking the invariant combination of the momenta, 
they are thus  expressed as 
\be
\label{masslessut}
    (\psi_L)^M_{-1/2} \vert 0; K_L \rangle  \otimes 
     (\psi_R)^\mu_{-1/2} \vert 0; K_R \rangle
    \comma
\ee
where $ \vert 0; K_{L/R} \rangle$ are the oscillator vacua with
the ten-dimensional momenta, 
\begin{align}
\label{KLR}
 &K_L^M = (k^\mu_L, p^a_L) \comma \quad
  K_R^M = (k^\mu_R, p^a_R) \comma 
\end{align}
whose components are moreover split into
\begin{align}
\label{kpLR}
 &  k^\mu_L= (k^{\tilde\mu}, k^{p}_L)
  \comma \quad
   k^\mu_R= (k^{\tilde\mu}, k^{p}_R)
   \comma 
 \end{align}
with $\tilde\mu=0, ..., 3$ and $p=4,5$.
For the conjugacy classes in \eqn{ut0},  
the momenta $p^a_{L/R}$
for $X^a_{L/R}$ take the value on the root lattice, 
\be
\label{kpLR2}
 p^a_{L/R} = \sqrt{2 \over \alpha'} \, e^a_i m_{L/R}^i  \ \ (m_{L/R}^i \in \bbZ)
 \period
\ee
The momenta along $X^{p}$ generally take the form, 
\be
\label{X45k}
   k_L^{p} = \frac{n_{p}}{R_{p}}
    + \frac{w^{p}R_{p}}{\alpha'}
   \comma \qquad
    k_R^{p} 
   = \frac{n_{p}}{R_{p}} - \frac{w^{p}R_{p}}{\alpha'}
   \comma
\ee
where 
$ n_{5} \in \bbZ/2 $ as in  \eqn{Ztw},
$ n_{4}, w^4, w^5 \in \bbZ $, 
and $R_4$ is the compactification radius for $X^4$. 
In this notation, 
$n_5 \in \bbZ$,  $w^5 \in 2 \bbZ$ for the states in \eqn{masslessut}.
Because of  the on-shell  condition, 
these states are 
indeed massless from the ten-dimensional point of view.
Together with the level-matching condition,  the momenta satisfy 
\be
\label{masslessuntw}
  K_L\cdot K_L := \eta_{MN} K_L^M K_{L}^N = 0
  \comma \quad
   K_R\cdot K_R := \eta_{MN} K_R^M K_{R}^N = 0 
   \period 
\ee
These strings are regarded as those obtained from
 the four-dimensional massless strings by exciting the momenta 
along $S^1 \times \bbR_{\rm base} \times T^4_{\rm fiber}$,
which make them  massive from the four-dimensional point of view.

The corresponding vertex operators
are constructed in a standard manner.
In the $(-1,-1)$ picture, they are given by%
\footnote{
We have omitted the cocycles for the toroidal compactification. 
They give extra phases in 
the amplitudes below, which are however irrelevant in our discussion.
}
\be
\label{vertexut0}
  V_{{\rm ut}; p_R}^{(-1,-1)} = g_c 
  e^{-\phi_L - \phi_R} 
  \zeta_M \psi_L^M \zetabar_\mu \psi_R^\mu \, 
  e^{i K_L \cdot X_L+i K_R \cdot X_R }
  \comma
\ee
where $g_c$ is the coupling constant, 
$\zeta_M, \zetabar_{\mu}$ represent the polarizations
satisfying $K_L^M \zeta_M = k_R^\mu \zetabar_\mu = 0$, and
$k_R\cdot X_R := \eta_{\mu\nu}k_R^\mu X_R^\nu $. 
$\phi_{L,R}$ are the bosonized superconformal ghosts with the normalization, 
$\phi_L(z) \phi_L(0) \sim - \ln z$, $\phi_R(\zbar) \phi_R(0) \sim - \ln \zbar$. 
We have explicitly denoted the dependence of $V_{{\rm ut}; p_R}^{(-1,-1)}$ 
on $p_R^a$.
Acting with the modes of the supercurrents,
one also obtains the vertex operators in  the $(0,0)$ picture,
\begin{align}
\label{V00}
   & V_{{\rm ut}; p_R}^{(0,0)}   \\
   &= -{2g_c \over \alpha'}
  \zeta_M \bigl(i \del X_L^M + \half \alpha' (K_L\cdot \psi_L)
   \psi_L^M \bigr) \, \zetabar_\mu
  \bigl(i \del X_R^\mu + \half \alpha' (K_R\cdot \psi_R) \psi_R^\mu \bigr)\, 
  e^{i K_L \cdot X_L+i K_R \cdot X_R } \period \nn
\end{align}

To make the states invariant under the twist $\sigma$, 
we further 
take the even combination of the states with
$p_R$ and $-p_R$. 
The corresponding vertex operator is denoted e.g. by
\be
\label{invvertex}
  V_{{\rm inv}; p_R}^{(0,0)} := \frac{1}{\sqrt{2}} \left( V_{{\rm ut}; p_R}^{(0,0)}
  + V_{{\rm ut}; -p_R}^{(0,0)}\right) \period
\ee

\subsection{Massless strings in the twisted sector}

\label{sec:twmassless}

For the twisted sector, 
the conjugacy classes in $F^{T^4}_{(1,m)}$  for the left-mover 
are specified by the momentum $p_L$ 
as in the untwisted sector. However, 
$p_R$  in the right-mover vanishes.
On dimensional grounds, one finds that 
${\rm \overline{o}/\overline{v}}$ stands for the states with
an even/odd number of excitations of $\tilde{\alpha}^a_{-(1/2+k)}$
$(k \in \bbZ_{>0})$, 
whereas ${\rm \overline{s}, \overline{c}}$ stand for those
corresponding to their excitations 
above the twist fields.
As in the untwisted sector, 
the invariant states are obtained 
in such a way that the signs due to the twist are cancelled with each other.
For  the twisted sector 
(of the internal $T_{\rm fiber}^4$ part),
we consider the ten-dimensional massless strings which belong to
the sector with $w \in 2 \bbZ+1 $ and $n \in 2 \bbZ$
in (\ref{Ztw}). The relevant part of the partition function reads
$  \big(F^{T^4}_{(1,0)}\overline{f_{(1,0)}}
 + 
  F^{T^4}_{(1,1)} \overline{f_{(1,1)}} \big) \calJ$.
We concentrate on the NS-NS sector 
with respect to the  fermions 
$\psi_L^M$ $(M=2, ..., 9)$ and $\psi_R^\mu $ $(\mu=2, ..., 5)$
without the twist.%
\footnote{
Though the NS and Ramond sectors
are mixed due to the twist, 
the periodicity of the matter supercurrent 
$T_{F;R}^m$ in the right-mover
does not change
since the boundary conditions of $X_R^a$ and $\psi_R^a$
are changed simultaneously. 
The NS/Ramond sector of the matter part
with respect to  
 $\psi_R^\mu $
thus couples to the NS/Ramond sector of the superconformal ghosts,
and thus the BRST symmetry is maintained. 
} 
The conjugacy class including the NS-NS massless
states is then labeled as 
\be
\label{twgr}
   {\rm \bigm[o\, \overline{(oc)}  \bigm]}_X \times  
    {\rm \bigm[v \, \overline{(os)} \bigm]}_\psi
\period
\ee
The left-moving part is the same as  in \eqn{ut0}.
The conjugacy class ${\rm \overline{c}}$
for $X^a_R$
should include, as mentioned above,
the states which correspond to
the twist field
$\Sigma_R(\zbar) = \prod_{a=6}^9\Sigma_a (\zbar)$ 
implementing 
the chiral reflection
 \eqn{crefX} and its dual $\bar\Sigma_R(\zbar)$, 
  and the excited ones above them.
Here, $\Sigma_a$ are the twist fields for 
$X_R^a$
 with dimension $1/16$, and regarded as the twist fields
 of the type 
 in the Ashkin-Teller model 
 \cite{Bershadsky:1986mt,Dixon:1986qv,Zamolodchikov:1987ae,
 Ginsparg:1988ui,Hashimoto:1996he,Frohlich:1999ss,Mukhopadhyay:2001ey,
 Mattiello:2018kue}. 
 In appendix \ref{app:ATtw},  we summarize the properties 
 of those twist fields which are used below. 
For $\psi_R^\mu$, the conjugacy class $\rm \overline{os}$ represents
the states corresponding to the twist fields $\hat{S}_R^\alpha$
with dimension $1/4$,
and the excited ones above them. 

The spectrum in the
partition functions $F^{T^4}_{(w,m)} $ in \eqn{Fab} 
can be represented by free fermions. In that case, the T-duality twist
on $X_R^a$ is represented similarly to \eqn{crefH}
by the corresponding free bosons through bosonization.
Such a representation would simplify 
the computations of scattering amplitudes. However,
those free bosons are related non-locally to the original 
bosons $X_R^a$, and thus the physical interpretation  of the
results is not clear as they stand. 
We thus do not take this route in the following.
The problem is essentially equivalent 
to the ``bosonization'' of the twist fields $\Sigma_a$.
For a recent discussion in this respect, 
see \cite{Mattiello:2018kue}.

In the $(-1,-1)$ picture, 
the corresponding vertex operators are then,
\be
\label{Vtw0}
  V_{\rm tw}^{(-1,-1)} 
  = g'_c 
  e^{-\phi_L - \phi_R} 
  \zeta_M \psi_L^M \, 
\ubar_\alpha \hat{S}_R^\alpha \, \Sigma_R \, 
  e^{i K_L \cdot X_L+i K_R \cdot X_R }
    \comma
\ee
or those with $\bar\Sigma_R$ instead of $\Sigma_R$, which we denote
by $\bar{V}_{\rm tw}^{(-1,-1)}$.
Here, $g_c'$ is the coupling 
and  $\ubar_\alpha$ is the polarization of the spinor. 
We denote the momenta 
$K_{L,R}$
again by the same form as in 
\eqn{KLR}-\eqn{X45k}. 
From the physical state conditions, one finds  that 
the corresponding states are 
massless from the ten-dimensional point of view, and 
\be
\label{masslesstw}
  K_L \cdot K_L =  K_R \cdot K_R = 0 \period
\ee
In the notation \eqn{X45k}, 
the momenta 
$k_{L,R}^5$
of these states have 
$n_5 \in \bbZ$, $ w^5 \in 2\bbZ +1$, 
whereas the momenta $p_L^a$ are on the root lattice and 
$p^a_{R}$ are vanishing due to  the twist, 
\be
\label{pLRtw0}
   p^a_L = \sqrt{2 \over \alpha'} \, e^a_i m_L^i  
   \ \ (m_L^i \in \bbZ)
   \comma  \qquad p^a_R = 0 \period
\ee
The strings which we are considering
are regarded as those obtained from
the ground states, or the six-dimensional massless states outside 
$T^4_{\rm fiber}$, in the twisted sector
 by exciting the internal momenta. 
They are massive from the four-dimensional point of view.

As in the case of symmetric orbifolds, 
each twisted sector and hence each twist field 
may be associated 
to a fixed point under the chiral reflection \eqn{crefX}. Thus, 
the vertex operators generally include the fields which implement
the shifts among the fixed points \cite{Hamidi:1986vh}. 
In the following, we work in a twisted sector with the same fixed point
for simplicity, and assume that the twist field $\Sigma_R$ is associated to it.%
\footnote{The conjugacy class c of so(4) has two states 
 with dimension 1/4. A possible identification would be that each corresponds
 to $\Sigma_R$ or $\bar\Sigma_R$.
 Another possibility would be that $\Sigma_R$ is self-dual 
 as in the Ashkin-Teller model, and
 one corresponds to $\Sigma_R$ and the other to 
 another twist field for another fixed point.
 The following discussion does not depend on details of the identification.
See appendix \ref{app:ATtw}.
For the $\bbZ_2$-symmetric orbifolds 
with Lie algebra lattices, 
the fixed points have been discussed in detail in \cite{Itoh:1987bz}.
 }
\section{Amplitudes with vanishing right-moving internal momenta}

Now, we are ready to compute the amplitudes of
the strings in the twisted and untwisted sectors 
whose vertex operators are constructed
in the previous section.
We focus on those of
the NS-NS states as mentioned above. 
The amplitudes involving the Ramond states are related by supersymmetry.
In particular, 
let us first consider the case where 
the right-moving internal
momenta $p_R$'s along $T^4_{\rm fiber}$ vanish
also in the untwisted sector, while $p_L$'s are kept generic both 
in the twisted and untwisted sectors.
This serves as  a preparation for 
the case with non-vanishing $p_R^a$
discussed in the next section, as well as a check of our computations there.

\subsection{Three-point amplitudes involving the twisted sector}
\label{sec:3pt}
The untwisted
NS-NS states in section \ref{sec:utmassless} \
are invariant under the twist $\sigma$ for the T-fold
 after taking 
the invariant combination of the momenta.
The correlation functions and amplitudes only among them
at the tree level are the same
as those in the original model without the twist. 

Once the twisted sector is involved, a non-vanishing 
three-point amplitude needs to include two twisted states,
as they change 
the Hilbert space of the untwisted sector
to that of the twisted sector, and vice versa.
To saturate the ghost charge, the amplitude takes the form,
\be
A_3^{(0)} = 
\Big\langle
 c\cbar V_{{\rm ut};p_R=0}^{(0,0)}(z_1)   \, 
 c\cbar \bar{V}_{\rm tw}^{(-1,-1)}  (z_2)
\, c\cbar V_{\rm tw}^{(-1,-1)}
(z_3) \Big\rangle  \comma
\ee
with
$c(z)$, $\cbar(\zbar)$ being the ghosts.
We note that $V_{{\rm ut};p_R=0} = \frac{1}{\sqrt{2}}V_{{\rm inv};p_R=0}$ 
for $p_R = 0$. Using the physical state conditions, one finds 
\begin{align}
\label{A31}
 A_3^{(0)} & = -(iC_{S^2}) 
 g_c g_c'^2 \, a_3^L \, a_3^R 
   (2\pi)^{10}\delta^{(10)}(K_1+K_2+K_3) 
 \comma \nn \\
 & a_3^L = - \sqrt{\frac{\alpha'}{2}} \zeta_{1M}\zeta_{2N}\zeta_{3K}
  t^{MNK} \comma \quad
  t^{MNK} := \eta^{MN} K_{L2}^K +  \eta^{NK} K_{L3}^M
   + \eta^{KM} K_{L1}^N  \comma \\
    & a_3^R = - \sqrt{\frac{\alpha'}{2}}
     \ubar_{23} 
     \zetabar_{1\mu} k_{R3}^\mu = 
    \frac{1}{2}\sqrt{\frac{\alpha'}{2}}
    \ubar_{23}\zetabar_{1\mu} (k_{R2}^\mu-k_{R3}^\mu) \comma
    \nn
\end{align}
where 
$\bar{u}_{ij} := \bar{u}_{i\alpha}  \calC^{\alpha\beta}\bar{u}_{j\beta}
= \ubar_{i\alpha} \ubar_{j}^\alpha$
with $\calC^{\alpha\beta}$ being (the chiral block of) the so(4) charge 
conjugation matrix, 
and a constant $iC_{S^2} $ comes from the determinant 
of the Laplacian. 
$K_{L/Rj}$ is the left/right-moving momentum 
for the $j$-th vertex.
Since $p_{Rj}^a =0$ in the present case, $K_{Rj}^M = (k_{Rj}^\mu, 0)$.
$\zeta_{jM}$ and $\zetabar_{j\mu}$ are the vector polarizations 
for the $j$-th vertex and,  similarly, 
$\bar{u}_{j\alpha}$ are the spinor polarizations. 
Regarding the factors representing the momentum conservation, 
we normalize the states for the internal part without the volume
factor as for  
the partition function \eqn{TfoldZ}, and thus 
 $(2\pi)\delta(K^p)$  for $S^1 $ or $\bbR_{\rm base}$  
 and 
  $(2\pi)^4 \prod_{a=6}^9\delta(K^a)$ for $T^4_{\rm fiber}$
 stand for 
 $\delta_{n_p,0} \delta_{w^p,0}$ $ (p = 4,5)$
 and 
  $\prod_j \delta_{m_{L}^j,0}$, respectively.
\subsection{Four-point amplitudes}
\label{sec:pR04pt}

Let us move on to the computation of the four-point amplitudes.
The number of the twisted states therein should be even,
as in the three-point amplitudes.
We thus consider 
\be
\label{A4tw3}
A_4^{(0)} = 
\Big\langle
 c\cbar V_{\rm ut;0}^{(0,0)}(z_1) 
  \int d^2 z_2 \, V_{\rm ut;0}^{(0,0)}(z_2) 
 \, c\cbar \bar{V}_{\rm tw}^{(-1,-1)}(z_3)  \, c\cbar V_{\rm tw}^{(-1,-1)}(z_4)
\Big\rangle 
\comma
\ee
where we have set $p_{R}^a = 0$ for the untwisted states, 
and hence $p_{Rj}^a = 0$ $(j=1, ..., 4)$.  
The computation of $A_4^{(0)}$ is straightforward though needs some algebra.
For definiteness and later use, we list the result in some detail.
First, by setting $z_1 = 1$, $z_2 = z$, $z_3 \to  \infty $ and $z_4 =0$,  one finds 
\begin{align}
\label{A4tw2}
  A_4^{(0)} & = 
  (iC_{S^2})(g_c g_c')^2 \times
  (2\pi)^{10}\delta^{(10)} \Big( {\textstyle \sum_{a=1}^4 K_a} \Big)
  \times I^{(0)}_4  \comma
\end{align} 
where
\begin{align}
\label{I40}
& 
I^{(0)}_4   = \int d^2 z \, D_{4;L}(z) D_{4;R}^{(0)}(\zbar) \comma  \\
& 
D_{4;L} = 
e^{-\pi i (1+\frac{\alpha'}{2} K_{L3} \cdot K_{L4})}
z^{\frac{\alpha'}{2} K_{L2} \cdot K_{L4}} 
(1-z)^{\frac{\alpha'}{2} K_{L1} \cdot K_{L2}}
\cdot\frac{\alpha'}{2}
\left(
 \frac{A}{(1-z)^2}
 + \frac{B}{1-z} 
 + \frac{C}{z}
\right)
\comma \nn \\
& 
D_{4;R}^{(0)}  = 
e^{+\pi i (1+\frac{\alpha'}{2} k_{R3} \cdot k_{R4})}
\zbar^{\frac{\alpha'}{2} k_{R2} \cdot k_{R4}} 
(1-\zbar)^{\frac{\alpha'}{2} k_{R1} \cdot k_{R2}}
\cdot\frac{\alpha'}{2}
\left(
  \frac{\bar{A}^{(0)}}{(1-\zbar)^2}
 + \frac{\bar{B}}{1-\zbar} 
 + \frac{\bar{C}}{\zbar}
\right)\comma \nn
\end{align}
with
\begin{align}
\label{ABC}
&A = \zeta_{12}\zeta_{34} \left( \frac{2}{\alpha'} - K_{L1} \cdot K_{L2} \right) 
\comma \nn \\
&B  =    (K_{L1} \cdot K_{L2}) (\zeta_{14} \zeta_{23}-\zeta_{13} \zeta_{24})
+ \zeta_{12} (\kappa_{32} \kappa_{41}- \kappa_{31}\kappa_{42})
+  \zeta_{34} (\kappa_{12} \kappa_{24}- \kappa_{14}\kappa_{21}) 
\nn \\
& \qquad 
-\zeta_{13} \kappa_{21}\kappa_{43}
+\zeta_{14} \kappa_{21}\kappa_{34}
+\zeta_{23} \kappa_{12}\kappa_{43}
-\zeta_{24} \kappa_{12}\kappa_{34} \comma \nn
\\
&C   =  - (K_{L1} \cdot K_{L2}) \zeta_{13} \zeta_{24}
+ \zeta_{13} (\kappa_{21} \kappa_{42}- \kappa_{24}\kappa_{41})
+  \zeta_{24} (\kappa_{12} \kappa_{31}- \kappa_{13}\kappa_{32})  \\
& \qquad 
-\zeta_{12} \kappa_{31}\kappa_{42}
+\zeta_{14} \kappa_{31}\kappa_{24}
+\zeta_{23} \kappa_{13}\kappa_{42}
-\zeta_{34} \kappa_{13}\kappa_{24} \comma
\nn \\
&\bar{A}^{(0)}
 = \zetabar_{12} \ubar_{34}
 \Big( \frac{2}{\alpha'} - k_{R1} \cdot k_{R2} \Big)
 \comma \quad
 \bar{B}= \ubar_{34} 
 (\kappabar_{12}\kappabar_{24} - \kappabar_{21}\kappabar_{14})
 \comma \quad 
 \bar{C} = - \ubar_{34} \kappabar_{13}\kappabar_{24} \period \nn
\end{align}
The dot $(\, \cdot \, )$ stands for the contraction by $\eta_{MN}$
in the left-mover, and that by $\eta_{\mu\nu}$ in the right-mover.
To lighten the notation, we have introduced
\be
 \zeta_{ij} := \zeta_{i}\cdot \zeta_j \comma \quad
  \zetabar_{ij} := \zetabar_{i}\cdot \zetabar_j \comma \quad
  \kappa_{ij} := \zeta_{i} \cdot K_{Lj} \comma \quad
  \kappabar_{ij} := \zetabar_{i} \cdot k_{Rj} \period
\ee

The above integral is evaluated by the formula \eqn{zint}
in appendix \ref{app:integal}, which results in
 \be
 \label{I402}
I_4^{(0)} =  (-1)^{n_s} 
I(\alpha,\beta; n_t,n_s) 
\cdot a_{4;L} \, a_{4;R}^{(0)} \comma
\ee
where 
\begin{align}
\label{Ihat}
I(\alpha,\beta; n_t,n_s) & = 2 \pi (-1)^{1+n_u} 
{ \frac{\alpha'}{4} s_L \cdot \frac{\alpha'}{4} t_L 
\cdot \frac{\alpha'}{4} s_R \cdot \frac{\alpha'}{4} t_R
\over
 \Big(1+ \frac{\alpha'}{4} u_L \Big)
  \Big(1+ \frac{\alpha'}{4} u_R \Big)
}
{ \Gamma\Big( -\frac{\alpha'}{4} s_R \Big)
\Gamma\Big( -\frac{\alpha'}{4} t_R \Big)
\Gamma\Big( -\frac{\alpha'}{4} u_R \Big)
\over
\Gamma\Big( 1+\frac{\alpha'}{4} s_L \Big)
\Gamma\Big( 1+\frac{\alpha'}{4} t_L \Big)
\Gamma\Big( 1+\frac{\alpha'}{4} u_L \Big)
}
\comma \nn \\
a_{4;L}  &= 
  2 \cdot \Big(1+ \frac{\alpha'}{4} u_L \Big)
  \left[ 
  \frac{u_L}{s_L \Big(\frac{4}{\alpha'}+ s_L \Big)} A
  - \frac{1}{s_L} B   - \frac{1}{t_L} C
  \right] \comma \\
a_{4;R}^{(0)}  &= 
  2 \ubar_{34} \cdot \Big(1+ \frac{\alpha'}{4} u_R \Big)
  \left[ 
  \frac{u_R}{s_R \Big(\frac{4}{\alpha'}+ s_R \Big)} \bar{A}^{(0)} 
  - \frac{1}{s_R} \bar{B}   - \frac{1}{t_R} \bar{C}
  \right] \comma \nn
\end{align}
with
\be
\label{alphabetaL}
\alpha = \alpha' K_{L2} \cdot K_{L4} 
= - \frac{\alpha'}{2} t_L
\comma \quad
\beta = \alpha' K_{L1} \cdot K_{L2} 
= - \frac{\alpha'}{2} s_L
\period
\ee
We have also defined
\begin{align}
\label{Mandelstam}
 &s_L:= - (K_{L1} + K_{L2})^2 \comma \quad
  t_L:= - (K_{L1} + K_{L3})^2 \comma \quad
   u_L:= - (K_{L1} + K_{L4})^2 \comma \nn \\
   &s_R:= - (k_{R1} + k_{R2})^2 \comma \quad
  \  t_R:= - (k_{R1} + k_{R3})^2 \comma \quad
  \ \ u_R:= - (k_{R1} + k_{R4})^2 \comma   \,
\end{align}
and
\be
\label{nstu}
  n_s := \frac{\alpha'}{4} (s_L - s_R) \comma \quad
  n_t := \frac{\alpha'}{4} (t_L - t_R) \comma \quad
  n_u := \frac{\alpha'}{4} (u_L - u_R) \period
\ee
From \eqn{kpLR2}, \eqn{X45k} and \eqn{pLRtw0}, 
these $n_s, n_t, n_u$ are integers.
The Mandelstam variables satisfy
\be
  s_L + t_L + u_L = 0 \comma \quad
  s_R + t_R + u_R = 0 \comma
\ee
which also implies $n_s+n_t+n_u = 0$.
Substituting these into \eqn{A4tw2} gives 
$A_4^{(0)}$.

By rescaling, the polarization part is rewritten as 
\begin{align}
\label{tildea4}
 \tilde{a}_{4;L}
 &:=  { \frac{\alpha'}{4} s_L \cdot \frac{\alpha'}{4} t_L 
\over
 \Big(1+ \frac{\alpha'}{4} u_L \Big)
} a_{4;L} \nn \\
&= \frac{\alpha'^2}{8} \biggr[\,
\frac{1}{2}   u_L s_L \zeta_{13}\zeta_{24}
- t _L\Big(
  \zeta_{12}\kappa_{32} \kappa_{41} 
+ \zeta_{14} \kappa_{21} \kappa_{34} 
 + \zeta_{23} \kappa_{12} \kappa_{43}
 + \zeta_{34} \kappa_{14} \kappa_{23} \Big) \\
 & \qquad \quad
 + (\mbox{terms with} \ 2 \leftrightarrow 3) 
 + (\mbox{terms with} \  3 \leftrightarrow 4)
 \,
 \biggr] \comma \nn \\
 \tilde{a}_{4;R}^{(0)}
 &:=  { \frac{\alpha'}{4} s_R \cdot \frac{\alpha'}{4} t_R 
\over
 \Big(1+ \frac{\alpha'}{4} u_R \Big)
} a_{4;R}^{(0)} 
 =  \frac{\alpha'^2}{8}
  \ubar_{34}
  \Big( \frac{1}{2}t_R u_R \zetabar_{12} 
 - t_R \kappabar_{14} \kappabar_{23} - u_R \kappabar_{13} \kappabar_{24} 
 \Big) \period \nn
\end{align} 
The left-moving part $\tilde{a}_{4;L}$ is 
symmetric with respect to $s_L, t_L, u_L$, 
and agrees with  
the standard expression for the massless scattering 
\cite{Green:1987sp,Blumenhagen:2013fgp}.
The amplitude is also expressed in a Kawai-Lewellen-Tye (KLT)-like form
\cite{Kawai:1985xq} through,
\begin{align}
I_4^{(0)} 
&= \tilde{a}_{4;L} \, \tilde{a}^{(0)}_{4;R} 
 \cdot  2(-1)^{n_s} 
 \sin\Big(\frac{\pi \alpha'}{4}  s_L\Big)
{ \Gamma\Big( -\frac{\alpha'}{4} s_L \Big)
\Gamma\Big( -\frac{\alpha'}{4} u_L \Big)
\over
\Gamma\Big( 1+\frac{\alpha'}{4} t_L \Big)
}
{ \Gamma\Big( -\frac{\alpha'}{4} s_R \Big)
\Gamma\Big( -\frac{\alpha'}{4} t_R \Big)
\over
\Gamma\Big( 1+\frac{\alpha'}{4} u_R \Big)
} 
\period
\end{align}

For comparison with the result in the next section, let us
note some properties of the amplitude. First, 
$A_4^{(0)}$ has the poles in the $s$-channel at
\be
  \frac{\alpha'}{4} s_L - n_s = \frac{\alpha'}{4} s_R = k 
  \quad (k \in \bbZ_{\geq 0}) 
\ee 
for $n_s \geq  0$, whereas at
\be
  \frac{\alpha'}{4} s_R - |n_s| = \frac{\alpha'}{4} s_L = k 
  \quad (k \in \bbZ_{\geq 0}) 
\ee 
for $n_s <  0$. The poles in the $t$- and $u$-channels are similar.
In the hard-scattering limit,
\be
 s_{L/R} \to \infty \comma \quad {t_{L/R} \over s_{L/R}} 
 : {\rm fixed}, 
\ee
with $n_s, n_t, n_u$ also fixed for simplicity,
one finds the asymptotic form,
\be
\label{hardscattA0LR}
 \log A_4^{(0)} \sim  \log A_{4;L}^{(0)} + \log A_{4;R}^{(0)} \comma
\ee
where $\log A_{4;L/R}^{(0)}$ comes from the left/right-mover
and is given by
\begin{align}
\label{hardscatt1}
  & \log A_{4;L/R}^{(0)} 
   \sim  -\frac{\alpha'}{4} 
  (s_{L/R} \log s_{L/R} 
  +t_{L/R} \log t_{L/R} 
  +u_{L/R} \log u_{L/R}  )
   \period
\end{align}
This behavior is obtained by evaluating 
the saddle point value of the integrand of $I_4^{(0)}$
\cite{Gross:1987ar}.
Although the Lorentz symmetry is absent for the internal part 
$K_{L/R}^M$ $(M=4, ..., 9)$, let us further 
suppose for simplicity that
e.g. the left-moving spatial momenta $\vec{K}_L = (K_L^M)$ $(M=1, ..., 9)$ 
satisfy
\be
\label{comKL}
0= \vec{K}_{L1}+ \vec{K}_{L2} = -(\vec{K}_{L3}+ \vec{K}_{L4})
\comma
\ee
for the scattering process $1+2 \to 3 + 4$, 
as in the usual center-of-mass frame.
In this case,%
\footnote{
We take
all the spacial momenta to be  positive when they are incoming.
}
\be
\label{stuLK}
  s_L = 4K^2 \comma \quad
  t_L = -2K^2 (1+ \cos \theta) \comma \quad
  u_L = -2K^2 (1- \cos \theta)  \comma
\ee
with $ K:= | \vec{K}_{Lj}|$ $(j=1, ..., 4)$. The left-moving part of 
\eqn{hardscatt1} then becomes
\be
\label{logA4L0}
 \log A_{4;L}^{(0)} \sim 
 -\alpha' K^2 f(\theta) \comma
\ee
where
$\theta$ is the angle between $\vec{K}_{L1}$ and $\vec{K}_{L3}$
defined via $\vec{K}_{L1} \cdot \vec{K}_{L3} = K^2 \cos \theta$,
and 
\be
\label{hsangle}
f(\theta) = 
 - \cos^2 \frac{\theta}{2} \cdot \log \Bigl(\cos^2 \frac{\theta}{2}\Bigr)
  -  \sin^2 \frac{\theta}{2} \cdot \log \Bigl(\sin^2 \frac{\theta}{2} \Bigr)
    \comma
\ee
as in the standard case. 
In the Regge limit,
\be
 s_{L/R} \to \infty \comma \quad t_{L/R}, n_s, n_t, n_u : {\rm fixed},
\ee
one also finds
\be
\label{Regge1}
    \log A_{4;L/R}^{(0)} 
   \sim  \frac{\alpha'}{4}  t_{L/R} \log s_{L/R}
  \period
\ee
In  the limit $|s_{L/R}|, |u_{L/R}| \gg |t_{L/R}|$, \eqn{hardscatt1} 
reduces to \eqn{Regge1}.

\section{Amplitudes with non-vanishing right-moving internal momenta}

Based on the discussion in the previous section,
we now consider the general amplitudes where 
the right-moving internal
momenta $p_{Rj}$ along $T^4_{\rm fiber}$
are non-vanishing in the untwisted sector, while $p_{Lj}$ are kept 
generic as before. 
The behavior of the amplitudes is indeed
 changed due to the T-duality twist.

\subsection{Three-point amplitudes}
\label{sec:3pt2}

When $p_R \neq 0$ in the untwisted sector, 
the three-point amplitude involving the twisted sector 
becomes
\be
A_3 = 
\Big\langle
 c\cbar V_{{\rm inv};p_{R1} \neq 0}^{(0,0)}(z_1) 
 \, c\cbar \bar{V}_{\rm tw}^{(-1,-1)}(z_2) 
\, c\cbar V_{\rm tw}^{(-1,-1)}
(z_3) \Big\rangle  \period
\ee
This is evaluated by using the correlator
\eqn{twistedcorr2} in appendix \ref{app:ATtw}. 
As discussed shortly, the coupling $g_c$ also comes to dependent 
on $p_R$, which we denote as $g_c(p_R)$. Up to the coupling part, 
the result of $A_3$ is then given by a combination of 
$A^{(0)}_3$ in \eqn{A31} weighted by
the phases  $e^{\pm i\bar{x}_0 \cdot p_{R1}}$,
\be
A_3 = \frac{1}{\sqrt{2}} (e^{ i\bar{x}_0 \cdot p_{R1}} 
+ e^{- i\bar{x}_0 \cdot p_{R1}}) \times \frac{g_c(p_R)}{g_c(0)} 
A^{(0)}_3 \comma
\ee
where
$\bar{x}_0 \cdot p_{R1} := 
\delta_{ab}\bar{x}_0^a \cdot p_{R1}^b$, and $\bar{x}_0^a$ is the zero-mode 
of the twisted sector which we are working in. 
As explained in the appendix, 
the momentum conservation is not imposed on $p_R$.

Supposed that the period of $X_R^a$ is half
that of $X^a = X_L^a+X_R^a$, which is given below \eqn{vielbein},
the periodicity of $X_R^a$ is represented as
$X_R^a \sim  X_R^a +\sqrt{\alpha'/ 2} \cdot  \pi e^a_k$
for each $k=6, ..., 9$. 
The fixed points under the twist $X_R^a \to - X_R^a$ are then at 
$\bar{x}_0^a  =  \sqrt{\alpha'/2}  \cdot (\pi/2) \sum_{k=6}^9 
\ep_k e^a_k $ where $\ep_k = 0,1$. 
From \eqn{kpLR2},
the phase $e^{2i\bar{x}_0 \cdot p_{R1}}$
becomes 
of the form $(-1)^m$ with $m \in \bbZ$ in this case.

\subsection{Four-point amplitudes} 
\label{sec:4pt2}

We now consider the four-point amplitude,
\be
\label{A4tw4}
A_4 = 
\Big\langle
 c\cbar V_{{\rm inv};p_{R1}}^{(0,0)}(z_1) 
  \int d^2 z_2 \, V_{{\rm inv};p_{R2}}^{(0,0)}(z_2)
 \, c\cbar \bar{V}_{\rm tw;0}^{(-1,-1)}(z_3)  \, c\cbar V_{\rm tw;0}^{(-1,-1)}(z_4)
\Big\rangle 
\comma
\ee
with $p_{R1}, p_{R2} \neq 0$. 
The computation of the left-moving part is the same as 
 in the previous section. For the right-moving part, 
by using the correlator in \eqn{twistedcorr3}
we first compute the amplitudes with $V_{{\rm ut};p_{Rj}}^{(0,0)}$ 
instead of $V_{{\rm inv};p_{Rj}}^{(0,0)} $ $(j=1,2)$, which we denote by
$A_4^{p_{R1},p_{R2}}$, and then sum up them.
The computation of $A_4^{p_{R1},p_{R2}}$ 
is similar to that in the previous case with $p_R=0$, 
but with some differences:
Firstly, $\big\bra \prod e^{ik_{Rj} \cdot X_R} \big\ket
\big\bra \bar\Sigma_R \Sigma_R \big\ket$ 
is replaced with
\begin{align}
\label{conteipx}
&\Big\bra \prod e^{ik_{Rj} \cdot X_R} \Big\ket
\Big\bra e^{ip_{R1} \cdot X_R} e^{ip_{R2} \cdot X_R} 
\bar\Sigma_R \Sigma_R \Big\ket
\nn \\
&  = \ 
e^{i\bar{x}_0 \cdot(p_{R1}+p_{R2})}
\times \zbar_{34}^{-{1 \over 2}}(\zbar_{12}\zbar_{34})^{-{\alpha' \over 4}s'_R}
(\zbar_{13}\zbar_{24})^{-{\alpha' \over 4}t_R}
(\zbar_{14}\zbar_{23})^{-{\alpha' \over 4}u_R}
 \left(\frac{1-\sqrt{\bar\xi}}{1+\sqrt{\bar\xi}}\right)^{\frac{\alpha'}{2}p_{R1}\cdot p_{R2}}
\comma
\end{align}
where $p_{R1} \cdot X_R := \delta_{ab} p_{R1}^a X_R^b$,
$p_{Ri} \cdot p_{Rj} := \delta_{ab} p_{Ri}^a  p_{Rj}^b$ and
 $\bar\xi :=\zbar_{13}\zbar_{24}/\zbar_{23}\zbar_{14} $.
 The last factor is understood as due to the twisted 
 propagator in \eqn{twprop}.
We have also defined
\be
  s'_R := s_R +k_{R1}^2 + k_{R2}^2 = -2k_{R1} \cdot k_{R2} 
  = -2k_{R3} \cdot k_{R4}- p_{R1}^2 - p_{R2}^2 \period
\ee
$s_R, t_R$ and $u_R$ are defined in \eqn{Mandelstam}. We note that e.g.
$t_R =  -2k_{R1} \cdot k_{R3} + p_{R1}^2$ due to $p_{R1} \neq 0$.
 The terms with $i\delbar X_R$ in the computation are changed similarly.

Secondly,  the terms 
$p_{Rj} \cdot \psi_R$ from $K_{Rj} \cdot \psi_R$ 
$(j=1,2)$ are not vanishing
 in the vertex $V_{{\rm ut}; p_R}^{(0,0)}$ in \eqn{V00}.
Consequently, one has  an extra term proportional to
\be
\Big\bra \psi_R^a(\zbar_1)  \psi_R^b(\zbar_2) 
  \hat{S}_R^\alpha(\zbar_3)\hat{S}_R^\beta(\zbar_4)\Big\ket
 = h_+(\zbar_j) \delta^{ab} \calC^{\alpha\beta} - h_-(\zbar_j) 
 (\calC\gamma^{ab})^{\alpha\beta}  \comma
\ee 
where the index structure on the right side is fixed by the symmetry.
We have defined 
$\gamma^{ab}:= \frac{1}{2}(\gamma^a\bar\gamma^b 
- \gamma^b\bar\gamma^a)$
with $\gamma^a$, $\bar\gamma^a$ 
being the chiral blocks of the so(4) gamma matrices.
Our conventions of those matrices are summarized in appendix 
\ref{app:so4gamma}.
By considering specific cases of the indices, one finds 
the coordinate dependent factors $h_\pm(\zbar_j)$  to be
\cite{Kostelecky:1986xg}
\be
h_\pm(\zbar_j) = 
\frac{1}{2 \zbar_{12}\zbar_{34}^{1 \over 2}}
 (\bar\xi^{1 \over 2} \pm \bar\xi^{-{1 \over 2}})
\period
\ee
This results in a change of the polarization factor,
\begin{align}
\label{Abar}
 \bar{A}^{(0)} \to &  \
 \bar{A}(\bar\xi) := \bar{A}^{(0)}
 - q_- \bar\xi^{1 \over 2} - q_+ \bar\xi^{-{1 \over 2}}
   \comma
\end{align}
with
\be
 q_\pm := \frac{1}{2} \zetabar_{12} \Big( \bar{u}_{34}(p_{R1} \cdot p_{R2})  
 \pm  p_{R1}^a p_{R2}^b  
 (\calC \gamma_{ab})^{\alpha\beta} \ubar_{3\alpha}\ubar_{4\beta} \Big)\period
\ee

After setting 
$\zbar_1 = 1$, $\zbar_2 = \zbar$, $\zbar_3 \to  \infty $ 
and $\zbar_4 = 0$, we find that $I_4^{(0)}$ in \eqn{I40} changes as 
\begin{align}
\label{I4}
I_4^{(0)} \ \to \ I_4   = \int d^2 z \, D_{4;L}(z) D_{4;R}(\zbar) \comma 
\end{align} 
for $A_4^{p_{R1},p_{R2}}$, 
where $D_{4;L}(z)$ is given in \eqn{I40} and 
\begin{align}
D_{4;R}  = &
e^{\pi i (1-\frac{\alpha'}{4} s'_R)}
\zbar^{-\frac{\alpha'}{4} t_R} 
(1-\zbar)^{-\frac{\alpha'}{4} s'_R}
 \left(\frac{1-\sqrt{\zbar}}{1+\sqrt{\zbar}}\right)^{\frac{\alpha'}{2}p_{R1}
 \cdot p_{R2}} 
 \frac{\alpha'}{2}
\left(
  \frac{\bar{A}(\zbar)}{(1-\zbar)^2}
 + \frac{\bar{B}}{1-\zbar} 
 + \frac{\bar{C}}{\zbar}
\right)\period
\end{align}
In the present case, the Mandelstam variables satisfy
\be
  s_L + t_L + u_L = 0 \comma \qquad
  s'_R + t_R + u_R = 0 \period 
\ee
$n_s, n_t, n_u$ defined in \eqn{nstu}
remain to be integers. We also define
\be
  n_s' := \frac{\alpha'}{4} (s_L -s'_R) 
  = n_s + \frac{\alpha'}{4}(p_{R1}^2 + p_{R2}^2) \comma
\ee
which is also an integer satisfying $n_s'+n_t + n_u = 0$.

The integral $I_4$ is evaluated by using the results in appendix \ref{app:integal}.
We thus obtain
\begin{align}
\label{A4twfin}
  A_4^{p_{R1},p_{R2}} & = 
  (iC_{S^2})
  (g_c')^2  g_c(p_{R1}) g_c(p_{R2})  \times
  (2\pi)^{10}\delta^{(10)} \Big( {\textstyle \sum_{a=1}^4 K_a} \Big)
  \times e^{i\bar{x}_0 \cdot(p_{R1}+p_{R2})} I_4  \comma \nn\\
   I_4  & =   2\pi (-1)^{1+n'_s} \, 
    \tilde{a}_{4;L} \times
   { 
\Gamma\Big( -\frac{\alpha'}{4} u_L \Big)
\over
\Gamma\Big( 1+\frac{\alpha'}{4} s_L \Big)
\Gamma\Big( 1+\frac{\alpha'}{4} t_L \Big) 
}
\times 
b_{4;R} \comma \\
b_{4;R}
&= 
\frac{\alpha'}{2}
\Bigl[\, 
J_R (\bar\alpha, \bar\beta-2, \bar\gamma) 
\cdot \bar{A}^{(0)}
- J_R \Big(\bar\alpha + \frac{1}{2}, \bar\beta-2, \bar\gamma \Big)  
\cdot q_-
- J_R \Big(\bar\alpha - \frac{1}{2}, \bar\beta-2, \bar\gamma \Big)  
\cdot q_+ 
 \nn \\
& \qquad \qquad 
 + J_R (\bar\alpha, \bar\beta-1, \bar\gamma)
 \cdot \bar{B}
 + J_R (\bar\alpha-1, \bar\beta, \bar\gamma)
 \cdot \bar{C} \, 
\Bigr] \comma \nn
\end{align}
where 
$J_R (\bar\alpha, \bar\beta, \bar\gamma) $
is given in terms of the hypergeometric function as in \eqn{I+result}, and
\be
\label{baralbega}
  \bar\alpha = -\frac{\alpha'}{4} t_R \comma \quad
   \bar\beta = -\frac{\alpha'}{4} s'_R \comma \quad
    \bar\gamma = \frac{\alpha'}{2} p_{R1} \cdot p_{R2} \period
\ee
We recall that
$\tilde{a}_{4;L}$ is given in \eqn{tildea4}, and 
$ \bar{A}^{(0)} $, $\bar{B}$, $\bar{C}$ are in \eqn{ABC}.
When $\bar\gamma=0$, the integral $I_4$ reduces to $I_4^{(0)}$ 
in \eqn{I402} due to  \eqn{I'gamma0}, 
and so does $A_4^{p_{R1},p_{R2}}$ to $A_4^{(0)}$.
We note that $A_4^{-p_{R1},-p_{R2}} = e^{-2i\bar{x}_0 \cdot(p_{R1}+p_{R2})}A_4^{p_{R1},p_{R2}}$.

By combining $A_4^{\pm p_{R1}, \pm p_{R2}}$, we obtain the 
four-point amplitude for the invariant untwisted states in \eqn{A4tw4},
\be
\label{invA4}
  A_4 = \frac{1}{2} \left(A_4^{p_{R1},p_{R2}}  + A_4^{p_{R1},-p_{R2}}
   +A_4^{-p_{R1},p_{R2}} + A_4^{-p_{R1},-p_{R2}}\right) \period
\ee
This is our final result of the computation of  $A_4$.

\subsection{Poles of the amplitudes}

From the analytic structure of $J_R (\bar\alpha, \bar\beta, \bar\gamma) $
explained in appendix \ref{app:integal}, 
one finds that 
the possible poles of
the amplitude $A_4$ are at 
\be
 \frac{\alpha'}{4} u_L = m_u \comma \quad
   \frac{\alpha'}{4} t_R = \frac{1}{2} (m_t+l_t) \comma 
   \quad
   \frac{\alpha'}{4} (s'_R -2 p_{R1} \cdot p_{R2} )= m_s + l_s \comma
\ee
where $ m_u, m_t, m_s \in \bbZ_{\geq 0}$,
and 
$l_t = 0,1,2,3$ and
$l_s = -1, 0, 1$ are due to the shift of the arguments
$\bar\alpha$ and $\bar\beta$ in $J_R$, respectively.
Since
$ s'_R  - 2 p_{R1} \cdot p_{R2}
 = -2 K_{R1} \cdot K_{R2} 
 = - (K_{R1} + K_{R2})^2 $, 
the poles are specified by the ``proper'' Mandelstam
variable also in  the $s$-channel.
From the integral representation 
of $I_4$ as $\zbar\to 1$, or 
by using \eqn{F4}, one can check that
the residue of the apparent unwanted tachyonic pole at  
$ s'_R  - 2 p_{R1} \cdot p_{R2} = -4/\alpha'$ from $l_s=-1$
vanishes. 
The series at $\frac{\alpha'}{4} t_R -\frac{1}{2} =m \in \bbZ_{\geq 0} $
should be cancelled with the zeros 
coming  e.g. from the hypergeometric
functions
in $I_4$
as in the case of $\bar\gamma =0$; 
the amplitude
is symmetric with respect to $t_R$ and $u_R$, or
the $t$-channel poles arise from the contributions around $\zbar \to 0$ 
which are irrelevant of
the value of 
$\bar\gamma$.
This agrees with the allowed intermediate states in this channel 
read off from the partition function \eqn{TfoldZ}.
Thus, we are left with the possible poles at 
\be
 \frac{\alpha'}{4} u_L = m'_u \comma \quad  \
  \frac{\alpha'}{4} t_R = m'_t  \comma  \quad  \   
    \frac{\alpha'}{4} (s'_R - 2 p_{R1} \cdot p_{R2})  
    =   m'_s \comma 
\ee
with $m'_u, m'_t, m'_s \in \bbZ_{\geq 0}$.
Still, some may be cancelled with the zeros from other part of $I_4$,
as in the case of $\bar\gamma =0$ described in section \ref{sec:pR04pt}. 
\subsection{Suppression of the three-point coupling} 

\begin{figure}[t]
   \centering
  \includegraphics[width=85mm]{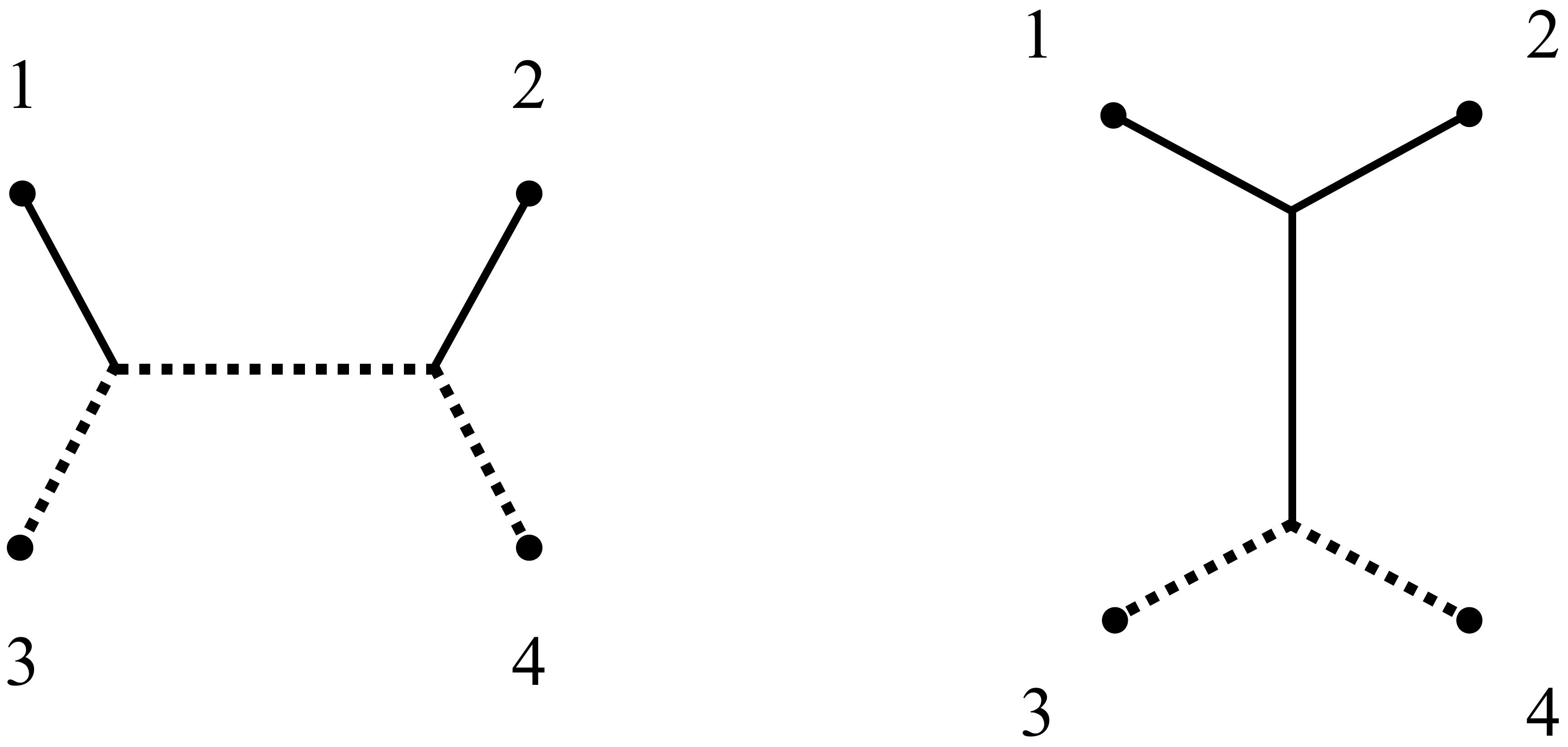} 
    \caption{
    Emission of untwisted strings from twisted sector or 
    $t$-channel picture of $A_4$ (left), and
    $s$-channel picture (right).
       The solid/dashed lines represent the propagation 
    of the untwisted/twisted strings. 
    }
    \label{fig:tschannel}
\end{figure}

In the presence of the two twist fields $\Sigma_R$, $\bar\Sigma_R$,
the untwisted vertices are regarded as describing the 
emission of the untwisted states from the twisted sector,
as in Figure \ref{fig:tschannel} (left).
The factorization of $A_4$ shows that the coupling $g_c$ in 
$V_{{\rm ut};p_R}$ comes to depend on the momentum,
as mentioned above, similarly to the case of
symmetric orbifolds \cite{Hamidi:1986vh}.

To see this, we note that the last factor in \eqn{conteipx} 
from the twisted propagator is 
reduced to $\big[ (1-\sqrt{\zbar})/(1+\sqrt{\zbar}) 
\big]^{\frac{\alpha'}{2}p_{R1}\cdot p_{R2}}$ by 
setting $(\zbar_1, \zbar_2, \zbar_3 , \zbar_4) = (1,\zbar,\infty, 0) $ as above.
Though the denominator $1 +\sqrt{\zbar}$ does not change 
the short-distance singularity, it gives an extra factor in 
$ [ \, (1-\zbar)/4 \, ]^{\frac{\alpha'}{2}p_{R1}\cdot p_{R2}}$
 for $\zbar \to 1$. 
 We denote
the coupling in the vertex for the emission from the untwisted sector
by $g_0$.
Then, evaluating the contribution to $A_4$ from
the integrand around $\zbar =1$, 
and relating it  
to the product of the three-point amplitudes in the $s$-channel
as in Figure \ref{fig:tschannel} (right), we find 
\be
\label{A4gg}
  A_4 \sim g_c(p_{R1}) \, g_c(p_{R2}) \, 2^{-\alpha' p_{R1}\cdot p_{R2}}
  \sim g_0 \, g_c(p_{R1} +p_{R2} ) \period
\ee
Here, we have  used $C_{S^2} \sim 1/g_0^2$ \cite{Polchinski:1998rq}.
$g_c(p_R)$ is thus 
regarded as the three-point coupling among two twisted 
states and an untwisted state with 
internal momentum $p_R$.
In the $t$-channel factorization, one indeed finds
$A_4 \sim  g_c(p_{R1}) \, g_c(p_{R2}) $. 
The  relation \eqn{A4gg} is solved by
\be
\label{gcsuppress}
  g_c(p_R) \sim 2^{-\frac{\alpha'}{2}p_R^2} \, g_0 \period
\ee
In the present case of an asymmetric orbifold, 
the coupling depends only on the 
right-moving momentum $p_R$. 
Moreover, it follows from 
\eqn{kpLR2} 
that
\be
 \frac{\alpha'}{2} p_R^2 = C_{ij} m_R^i m_R^j \in 2 \bbZ \comma
\ee
where $C_{ij}$ is the Cartan matrix of so(8), 
and 
hence
the suppression factor of the coupling is 
quantized in integer powers of $1/4$.
The momentum along a compactified direction with radius $R_i$
has the winding contribution 
$w R_i/\alpha'$
 as in \eqn{X45k}, and $R_i/\sqrt{\alpha'}$ corresponds to the
 inverse coupling of the sigma model. Therefore, 
 the factor $2^{-\alpha'p_R^2/2}$
 in \eqn{gcsuppress} includes stringy effects 
 of the form $e^{- w^2 R_i^2/\alpha'}$ which are
non-perturbative from the sigma-model point of view.
The compactification radii in \eqn{pLR} 
 are of order $\sqrt{\alpha'}$ and not explicit there.
 This suppression of the coupling
implies that 
an untwisted string with large 
momentum $p_R$ 
hardly interacts with twisted strings confined around the fixed point.

The suppression factor in \eqn{gcsuppress}
is also understood as a consequence 
of the normal ordering of the vertex operator $e^{ipX}$, and of the
difference of the mode expansions for different boundary conditions by 
$n \in \bbZ$ or $n \in \bbZ + \frac{1}{2}$ \cite{Hamidi:1986vh}.
The mechanism of the suppression is hence robust 
even for asymmetric orbifolds. 
The argument of the normal ordering is extended to the case of 
$\bbZ_{N}$ orbifolds with $N > 2$.  
The suppression factor for  $\bbZ_{N>2}$ symmetric 
orbifolds
 schematically takes
the form $N^{-({\rm momenta})^2}  \times$
(factors depending on $N$ and momenta)
\cite{Hamidi:1986vh,Erler:1991an}. It is thus understood that
the base 2 of $1/4 = 2^{-2}$ in our case comes from $\bbZ_{N=2}$
of the T-duality twist in the untwisted sector
of the $T^4_{\rm fiber}$ part, 
and the exponent 2 is from the length of the 
roots of so(8).
The geometrical interpretation of the suppression 
 in the symmetric case\cite{Hamidi:1986vh}, however,  
may not apply as it stands,
since  we are considering a T-fold/asymmetric orbifold 
and in addition working in a twisted sector with the same fixed point.
\subsection{High energy behaviors} 
\label{subsec:HEbehavior}

From the expression \eqn{A4twfin}, \eqn{invA4}, 
 one can read off
high energy behaviors of the amplitude $A_4$.
The high energy limit corresponds to $\alpha' \to \infty$,
which is opposite to the particle limit  $\alpha' \to 0$.
To see those behaviors, let us first summarize the kinematics. 
We denote the momenta as in 
\eqn{KLR}-\eqn{X45k}, \eqn{pLRtw0}.
Their conservation reads
$0=  \sum_{j=1}^4 k_j^{\tilde\mu} 
    = \sum_{j=1}^4 k_{L/Rj}^{p} 
   =   \sum_{j=1}^4 p_{Lj}^{a} $, 
whereas
$p^a_{Rj}$ with $p^a_{R3}= p^a_{R4} =0 $ do not have to be conserved.
The momenta  also have to satisfy the on-shell and level-matching conditions
\eqn{masslessuntw}, \eqn{masslesstw}.
We mainly consider the scattering process $1 + 2 \to 3 + 4$.
Other processes can be analyzed similarly, 
which we comment on below.
By the Lorentz symmetry for $k^{\tilde\mu}$, one can then take 
the center-of-mass frame
where
\be
  0 = \vec{k}_1 + \vec{k}_2 = - (\vec{k}_3 + \vec{k}_4) \comma
\ee
with $ \vec{k} = (k^1, k^2, k^3) $. 
A variety of high energy limits is allowed because of
the asymmetry between the left- and the right-mover, 
the absence of the conservation of $p_R$ and that of the Lorentz symmetry 
for $k_{L/R}^p$ and $p_{L/R}^a$
along $S^1 \times \bbR_{\rm base} \times T^4_{\rm fiber}$. 
In the following, 
we concentrate on the case characteristic to 
the strings on the T-fold where $p_{R1}, p_{R2}$ are large
and the T-duality twist has significant influence
through the factor from the twisted propagator in \eqn{conteipx}.

\par\bigskip
\ni
{\bf Hard-scattering limit}

As an example, we first consider a high-energy limit where the momenta 
of the untwisted 
strings along $T^4_{\rm fiber}$
are large and those along other 
spatial directions are kept fixed:
\be
  | p_{L/R1} | \comma \ | p_{L/R 2} | \ \  \gg \ \ 
  |\vec{k}_j | \comma \ 
 | k^p_{L/Rj}| \quad (j=1,2)
   \period
\ee
We also fix the internal momenta of the twisted strings
$|k_{L/Rj}^p|$ $(j=3,4)$ for simplicity. 
From the on-shell and level-matching conditions, it follows that
$|p_{Lj}|$ $(j=3,4)$ are  fixed, and that 
\begin{align}
  |p_{Rj}| \sim  |p_{Lj}| \sim p \ \ (j=1,2)\comma \quad \ 
    | k_j^0|  \sim p \ \ (j=1,...,4)
   \comma \quad \ 
  |\vec{k}_j| \sim p  \ \ (j=3,4)
  \comma
\end{align}
with $p:= |p_{R1}| $ being large.
For this kinematic configuration, the Mandelstam variables are
\begin{align}
  &\ s_L  
 \sim 4p^2 
  \comma  \qquad \  
  t_L  
  \sim -2p^2
  \comma \qquad \ 
  u_L  
   \sim -2p^2 \comma  \nn \\
  & \  s'_R 
   \sim 2p^2 
  \comma \qquad \ 
  t_R 
   \sim - p^2
  \comma \qquad \ 
  \  u_R 
   \sim -p^2 \period 
\end{align}
Thus, sending $p \to \infty $ is a hard-scattering limit.

In this limit, the asymptotic form of the amplitude consists of
the terms from the left- and the right-mover as in  \eqn{hardscattA0LR},
and from the momentum-dependent couplings  in this case,
\be
\label{A4p12asym}
\log A^{p_{R1},p_{R2}}_4 \sim  \log A_{4;L} + \log A^{p_{R1},p_{R2}}_{4;R} 
+\log \bigl[g_c(p_{R1}) g_c(p_{R2})\bigr]
\period
\ee
The contribution from the left-mover 
comes from the factor $B^{-1}(-\alpha/2, -\beta/2)$ 
in \eqn{A4twfin}
with \eqn{alphabetaL}, where $B(x,y)$ is the beta function,
and remains the same as to 
$A_4^{(0)}$,
\be
\label{logA4L}
   \log A_{4;L} \sim 
   -\frac{\alpha'}{4} 
  (s_{L} \log s_{L} 
  +t_{L} \log t_{L} 
  +u_{L} \log u_{L}  ) 
  \sim - \alpha' p^2 \log 2 \period
\ee
Due to \eqn{gcsuppress}, the contribution from the couplings
 also becomes
\be
\log \bigl[g_c(p_{R1}) g_c(p_{R2})\bigr] \sim - \alpha' p^2 \log 2
\period
\ee

On the other hand, 
the contribution from the right-mover comes from 
the factor $J_R(\bar\alpha, \bar\beta, \bar\gamma)$ 
in \eqn{A4twfin}
with \eqn{baralbega} and
\be
 \bar\gamma
    \sim  \frac{\alpha'}{2} p^2 \cos \phi \comma
\ee
 where $\phi$ is the angle between $p^a_{R1}$ and $p^a_{R2}$.
The behavior of this factor for $p \to \infty$ 
can be found by using the formulas of the
hypergeometric function listed in appendix \ref{app:integal}.
Using \eqn{F1} and \eqn{F2}, one finds
\begin{align}
\label{I+HS}
  J_R(\bar\alpha, \bar\beta, \bar\gamma)  
  & \sim 2 B(2-\bar\beta, 1+\bar\beta + \gamma)
  F(2-\bar\beta, \bar\gamma-\bar\beta; 3+ \bar\gamma; -1) \nn \\
  &= 2^{\bar\beta-1}B(2-\bar\beta, 1+\bar\beta + \gamma)
  F(2-\bar\beta, 3+\bar\beta; 3+ \bar\gamma; 1/2)   \\
  & \sim \Big(\frac{\bar\beta}{2}\Big)^{-\bar\beta}
  \bigl(\bar\beta+\bar\gamma\bigr)^{\bar\beta+\bar\gamma \over 2}
   \bigl(\bar\beta-\bar\gamma\bigr)^{\bar\beta-\bar\gamma \over 2}
    \period  \nn 
\end{align}
Here, we have used 
\be
\label{barabc}
\bar\alpha \sim -\frac{1}{2} \bar\beta \comma \qquad
\bar\gamma \sim - \bar\beta \cos \phi \comma 
\ee
and 
the fact that $F(a,b;c;z)$ is 
an entire function 
of $a,b,c$ for 
$|z| < 1$ or  $\re z < 1/2$
and hence 
$
F(2-\bar\beta, 3+\bar\beta;3+\bar\gamma; 1/2) \sim 
F(-\bar\beta, 1+\bar\beta;\bar\gamma; 1/2) 
$ 
for large $|\bar\beta|$. 
Thus, when \eqn{barabc} holds,
\begin{align}
\label{logA4R}
 \log A_{4;R}^{p_{R1},p_{R2}}  \sim & -\frac{\alpha'}{4} 
  (s'_{R} \log s'_{R} 
  +t_{R} \log t_{R} 
  +u_{R} \log u_{R} )  -\frac{\alpha'}{4} s'_{R} \bigl[ \log 2 - f(\phi) \bigr] \nn  \\
  \sim & -\frac{\alpha'}{2} p^2  \bigl[ 2\log 2 - f(\phi) \bigr] \comma
 \end{align}
where $f(x)$ is the function defined in \eqn{hsangle}.
This function is rewritten as 
$f(\phi) = \tilde{f}\bigl( \cos^2 \frac{\phi}{2}\bigr)$ where
$\tilde{f}(y):=-y \log y - (1-y) \log(1-y)$. Since 
$ 0 \leq \tilde{f}(y) \leq \log 2$
for $y \in [0,1]$ with the maximum $\tilde{f}(1/2) = \log 2$ and the minima 
$\tilde{f}(0) = \tilde{f}(1) = 0$, 
one finds a bound,
\be
  4^{-\frac{\alpha'}{2} p^2}  \, \lesssim \,  A_{4;R}^{p_{R1},p_{R2}} 
  \,  \lesssim \,
 2^{ -\frac{\alpha'}{2} p^2} \period
\ee
When $\bar\gamma = 0$ or $\cos \phi = 0 $, namely,
when 
the factor from
the twisted propagator in \eqn{conteipx} disappears, 
$f(\phi)$
has the maximum $\log 2$ and the right-hand side of
\eqn{logA4R} reduces to the standard form in terms of 
$s'_R, t_R$ and $u_R$ as in \eqn{hardscatt1}. 
The angle dependence given by $f(\phi)$ is also the same 
as in \eqn{logA4L0}.
The sign of its coefficient $+\alpha' s'_R/4$ is, however, opposite
to \eqn{logA4L0}, so that the amplitude is more suppressed
as $|\bar\gamma|$ becomes larger.
The behaviors $A_{4;R}^{p_{R1},p_{R2}} \sim  2^{-\alpha' p^2/2}, 
 4^{-\alpha' p^2/2}$ for 
$\cos \phi=0$ ($\bar\gamma=0$), 
 $\cos \phi=\mp 1$ ($\bar\gamma=\pm \bar\beta$), respectively,
are also confirmed from  \eqn{I+reduce}, \eqn{F3} and \eqn{F4},
respectively.

The high-energy behavior in \eqn{I+HS} is obtained also 
 by   the saddle point $v_0$ 
of the integrand of  $J_R$ in \eqn{I+result}, which solves 
$(\bar\alpha +\bar\beta) v^2 + \bar\gamma v - \bar\alpha = 0 $.
For the present kinematic configuration, the solutions are
$ v_0 \sim -y \pm i \sqrt{1-y^2}$ with 
$y := {\bar\gamma}/{\bar\beta} \sim -\cos \phi$.
One finds that the two saddle points give the same 
value of the integrand up to a phase depending on $\bar\gamma$, 
and that it agrees with \eqn{I+HS}. 
This means that the saddle-point argument in \cite{Gross:1987ar}
may be developed also in our case of a T-fold.
Our explicit computation shows in particular that 
both of the two saddle points are dominant, 
and one can take either of them up to the phase 
corresponding to the choice of the branch, but not their linear combinations.

We note that \eqn{logA4R} in this limit is 
symmetric with respect to 
$\phi$ and $\pi - \phi$ or $\bar\gamma$ and $-\bar\gamma$
before summing up $\pm p_{Rj}$  as in \eqn{invvertex}
and \eqn{invA4}. Therefore,
in the asymptotic form 
of the total amplitude $A_4$ denoted as in \eqn{A4p12asym}, 
the contribution from the right-mover $A_{4;R}$ corresponding to 
$A_{4;R}^{p_{R1},p_{R2}}$ remains the same in this limit,
\be
  A_{4;R}  \sim  
    A_{4;R}^{p_{R1},p_{R2}}   \period 
\ee
With the help of the above saddle-point argument, 
this may be understood as follows:
If we formulate our orbifold theory 
on the Riemann surface with the branch cut representing
the twist, 
the summation of $\pm p_{Rj}$  is automatically taken care of
(up to the zero-mode factors $e^{i\bar{x}_0 \cdot p_{Rj}}$),
since moving to another sheet, $\sqrt{\zbar} \to - \sqrt{\zbar}$,  flips the sign of 
$\bar\gamma$ in \eqn{conteipx}. 
In picking up a saddle point, global properties of 
the Riemann surface are not relevant, and thus it may reproduces
the invariant result under $\bar\gamma \to - \bar\gamma$ 
or $\phi \to \pi - \phi$.
In the usual hard scattering, this type of invariance corresponds
to the symmetry between the $t$- and  $u$-channels, 
which is represented e.g. by $f(\theta)$ in  \eqn{hardscatt1}.
This observation may partly explain why the same angle dependence
$f(\phi)$ appears also in the present case.

It is also possible to consider the high-energy limit where
$|\vec{k}_1| = |\vec{k}_2| $ becomes large in addition. 
The asymptotic form of the amplitude
in this case would be obtained once the asymptotic form
of the hypergeometric function  in the general case is given,
or it is supposed that picking up one saddle point gives the correct 
result as above.

For other processes, one can consider, for example, 
the $1+3 \to 2 +4$ process where 
$ | p_{Lj} | \sim | p_{Rj} | \sim p$ $(j=1,2)$ are large and
other momenta are fixed or vanishing. 
The contribution from the right-mover and that from 
 the momentum-dependent couplings are dominant in this case.
 From the asymptotic form of $J_R$ as above, it turns out that, 
as $p$ increases, so does the former as opposed to \eqn{logA4R},  
while the latter still decreases. Adding these together, one  finds
$\log A_4 \sim - \frac{\alpha'}{2} p^2 f(\phi)$, which takes a standard form
for the hard scattering. This shows that the momentum dependence 
of the coupling ensures the soft behavior of the amplitudes in this case.

\par\bigskip
\ni
{\bf Regge-like limits}

Next, we consider high-energy limits of the $1+2 \to 3 +4$ process
where all the Mandelstam variables
do not become large. 
For simplicity, we suppose that 
$\vec{K}_{Lj} $ with $p_{Lj} \neq 0$ also 
satisfy the same relations  \eqn{comKL} and \eqn{stuLK} 
as in the usual center-of-mass frame. Then, 
\be
\label{LRegge}
  s_L \to \infty \comma \qquad  t_L : {\rm fixed}  \ \,  (\cos \theta \to -1) \comma
\ee
is a Regge(-like) limit. The contribution to the asymptotic form 
of $A_4$ from the left-mover 
takes the same form as $A^{(0)}_L$ in 
\eqn{Regge1}.
The Mandelstam variables in the right-mover are 
\begin{align}
 & s'_R 
 = 2 K^2 + 2k^2 - 2k_{R1}^p k_{R2}^p \comma \qquad
  t_R 
 = -2 K^2 + 2k^2 - 2k_{R1}^p k_{R3}^p + p_{R1}^2 \comma
\end{align}
where we have set $K=|\vec{K}_{Lj}|$ and $k=|\vec{k}_{Lj}|$ $(j=1, ..., 4)$.
To be concrete, we  take e.g. $p_L:= |p_{Lj}|$ $(j=1, ..., 4)$ to be large,
and $(k_{Lj}^p)^2$ $(j=1, ..., 4)$ and $(k_{Rj}^p)^2$ $(j=1,2)$ 
to be fixed.  
In this case, $K^2 \sim k^2 + p_L^2$ and, from
the level-matching condition, 
$ p_L^2
\sim p_{R1}^2 \sim p_{R2}^2 \sim (k_{R3}^p)^2 \sim
(k_{R4}^p)^2$, which results in%
\footnote{
For generic radii $R_p$ $(p=4,5)$ in 
\eqn{X45k}
where
$n_p$ and $w^p$ are separately conserved, 
\eqn{comKL} and \eqn{LRegge} imply 
$ {k}^p_{L1} ={k}^p_{L4} = -{k}^p_{L2} = -{k}^p_{L3} $
and a similar relation for  $k^p_{Rj}$.
The latter relation for the right-mover, however,  do not necessarily 
hold at special radii.
} 
\be
s'_R \sim 2(2k^2 + p_L^2) \comma \qquad
 t_R \sim - p_L^2
 \comma \qquad
 p_{R1} \cdot p_{R2} \sim p_L^2
 \cos \phi
 \period
\ee
With these Mandelstam variables, 
the analysis of the contribution 
from the right-mover  reduces to that for the
hard-scattering limit discussed above.

One may also consider the high-energy limit 
where $t_R$ instead of $t_L$ is fixed. 
Though $t_R$ is a fundamental variable of the amplitude,
it is not a standard Mandelstam variable due to the existence of
the internal momenta $p_{Rj}$. One finds 
that it is very restrictive to fix $t_R$ for large $p_{Rj}$
so as to be compatible with 
the momentum conservation, and the on-shell and 
level-matching conditions. We refrain from going into further details 
in such rather special cases.

\section{Discussion}

By taking a model of strings on T-folds as an example, 
we have discussed their interactions 
from the world-sheet point of view which are exact in $\alpha'$.
We have computed the three- and four-point amplitudes
of a class of ten-dimensional massless strings both 
in the twisted and untwisted sectors.
The four-point amplitudes are obtained in a closed form in terms of the hypergeometric function. From their factorization, 
it turns out that
the three-point coupling among the twisted and untwisted strings is 
 suppressed by the chiral momenta flowing along the T-folded
 internal torus, as in the case of general symmetric orbifolds.
 Furthermore, the coupling  
 is quantized in integer powers of 1/4 in our case of a T-fold, since
 the T-duality twist is $\bbZ_2$  in the untwisted sector 
 of the torus part and 
 the chiral momenta take the value on a Lie algebra lattice.
 The asymptotic forms of the four-point amplitudes in high-energy limits
have also been found. These results include stringy effects which are
non-perturbative from the sigma-model point of view, or
those for $\alpha' \to \infty$
which is opposite to the particle limit  $\alpha' \to 0$.
They probe strings on a T-fold beyond
the regime of supergravity and DFT.

For the strings other than those discussed in this paper, 
their vertex operators may be found
by following the discussion in section \ref{sec.vertexop},
and their amplitudes may be obtained similarly.
The momenta $p_{Rj}$ along the internal torus generally take the value 
on the weight lattice, instead of the root lattice, of so(8), and hence
the suppression factor of the coupling becomes integer powers of $1/2$,
instead of $1/4$.
By using the correlators involving more twist fields than those listed 
in appendix \ref{app:ATtw}, one can 
also compute 
the amplitudes with more twisted strings.
The simplest among these is 
the four-point amplitudes only of 
the strings excited above the ground states in the twisted sector
by the internal momenta.
In this case, however, the right-moving 
momenta $p_{Rj}$ along the internal torus, 
a key ingredient of our analysis, are all vanishing. 
The amplitudes involving different twisted sectors may be  
computed
by introducing the operators connecting them, 
as mentioned at the end of section \ref{sec:twmassless}.

As understood from the discussions so far,
our analysis relies only on general properties 
of the T-duality twist and the symmetry enhancement of the internal torus, 
and hence may be extended qualitatively to more general T-folds.
Furthermore,
our analysis can be applied or extended 
to other asymmetric as well as 
symmetric orbifold models, including those mentioned 
in section  \ref{sec:introduction} 
\cite{Satoh:2015nlc,Satoh:2016izo,Sugawara:2016lpa}, which 
are based on the (chiral) reflection of the coordinate fields.

Compared with toroidally compactified models, where
the left- and right-movers
are not symmetric either,
the appearance of the twisted propagator is characteristic
to the present case 
of a T-fold or an asymmetric orbifold. 
This changes the expression of the amplitudes and results
in the suppression of the coupling. 
The absence of the conservation 
of the momentum
$p_R$ along 
the twisted torus also affects the kinematics. Such properties
are common to the case of symmetric orbifolds.
Details are, however, different. For example, 
by extending the formula \eqn{I'result} as in \eqn{calJ},
the four-point amplitudes may be obtained as
a combination of the hypergeometric
functions from the left and right movers.
The exponent of the suppression factor in the symmetric case
is given by the momenta in the untwisted sector as
$p_L^2 + p_R^2$ \cite{Hamidi:1986vh} instead of $p_R^2$. Since
 $p_L$ and $p_R$ differ by the root lattice
as mentioned in section \ref{subsec:Tfoldso8}, 
these exponents can be largely different due to the windings $w^j$. 
As the left-mover is twisted in addition, 
the momentum $p_L$ in the twisted sector vanishes, 
and its conservation in the untwisted sector does not need to hold.
The left-moving part of the amplitudes 
may be given also by the variables which are defined 
similarly to $s'_R, t_R, u_R$  in the right-mover.
These change
the kinematics significantly and, together with the difference of
the spectrum,  the amplitudes 
with large left-right asymmetry
of the type discussed in section \ref{subsec:HEbehavior}  may not appear. 
The analysis of the high-energy behavior of the amplitudes
becomes different accordingly.

Our analysis in this paper demonstrates
that strings on T-folds, 
which are very stringy and might appear to be unconventional, 
can be analyzed 
in a quantitative manner, 
once the world-sheet theory is properly given, 
even though its construction is rather involved.
This may be regarded as an 
advantage of the world-sheet analysis.
A future problem there would be
to make  ``geometric'' interpretations clearer
regarding the fixed points, the associated twisted sectors, 
and the mechanism of the suppression 
of the coupling even for the non-geometric case of 
T-folds/asymmetric orbifolds.

Given the results which are valid for all $\alpha'$,  
one may consider their applications to further studies of strings on T-folds
and related ones. For example, 
since T-folds may be treated geometrically in DFT
before it is reduced in the supergravity frame, 
it would be of interest to figure out the implications 
of our results in the structure of DFT including 
the $\alpha'$-corrections \cite{Hohm:2014xsa}
 before the section condition is imposed. 
 In addition, non-geometric fluxes
 are associated with weakly constrained DFT \cite{Aldazabal:2013sca}.
 Thus, our results, or their extensions, would be used also to study its structure.%
 \footnote{
 We would like to thank the referee for raising a question on this issue.
 }
Finally, the world-sheet for 
the four-point amplitudes discussed in this paper has
the branch cut which is created by the twist fields
and implements the T-duality twist.
This branch cut may be regarded as a 
world-sheet conformal interface/defect
\cite{Wong:1994np,Petkova:2000ip,Bachas:2001vj}
 inducing T-duality. 
Conformal interfaces 
 have  properties which extend those of conformal boundaries
or D-branes. 
Although there are related works 
\cite{Bachas:2007td,Satoh:2011an,Bachas:2012bj,Elitzur:2013ut,Satoh:2015uia,Kojita:2016jwe},
the role of world-sheet conformal 
interfaces in string theory, if any, is yet to be explored. 
It would be interesting if one could probe it
by utilizing the results in this paper.
We would like to  discuss these issues further elsewhere.

\vspace{3ex}
\begin{center}
{\large\bf Acknowledgments}
\end{center}

We would like to thank Y. Sakatani and S. Watamura for useful discussion.
This work is supported in part by 
JSPS KAKENHI Grant Number JP17K05406.


\appendix
\renewcommand{\theequation}{\Alph{section}.\arabic{equation}}

\vspace{6ex}

\section{Components of partition function}
\label{app:partfn}

The components of the partition function in \eqn{TfoldZ}
are given by
 \begin{eqnarray}
 \label{ZRb}
 Z_{(w,m)}^{\rm base}(\tau,\bar{\tau}) &: =& 
\frac{\Rb}{\sqrt{\alpha'\tau_2}|\eta(\tau)|^2} 
e^{-\frac{\pi \Rb^2}{\alpha'\tau_2}|w\tau+m|^2} \comma 
\end{eqnarray}
\begin{equation}
\calJ(\tau) := \frac{1}{2} \left[ \left(\frac{\theta_3}{\eta}\right)^4
- \left(\frac{\theta_4}{\eta}\right)^4 - \left(\frac{\theta_2}{\eta}\right)^4 \right]
\equiv 0 \comma
\end{equation}  
\begin{align}
\label{FabT4}
& F^{T^4}_{(w,m)}(\tau,\bar{\tau})   \\ 
&  := 
\frac{1}{2} \cdot \left\{
\begin{array}{l}
e^{\frac{\pi i}{2} w}
\left\{
\left(\frac{\theta_3}{\eta}\right)^4
 \overline{\left(\frac{\theta_3\theta_4}{\eta^2}\right)^2}
+
\left(\frac{\theta_4}{\eta}\right)^4 
 \overline{\left(\frac{\theta_4\theta_3}{\eta^2}\right)^2}
+\left(\frac{\theta_2}{\eta}\right)^4
\overline{\left(\frac{\theta_2\vartheta_1}{\eta^2}\right)^2}
+
\left(\frac{\vartheta_1}{\eta}\right)^4
 \overline{\left(\frac{\vartheta_1\theta_2}{\eta^2}\right)^2} 
\right\} \comma  \\
e^{-\frac{\pi i}{2} m}
\left\{
\left(\frac{\theta_3}{\eta}\right)^4 
 \overline{\left(\frac{\theta_3\theta_2}{\eta^2}\right)^2}
+  \left(\frac{\theta_2}{\eta}\right)^4
\overline{\left(\frac{\theta_2\theta_3}{\eta^2}\right)^2}
-\left(\frac{\theta_4}{\eta}\right)^4 
\overline{\left(\frac{\theta_4\vartheta_1}{\eta^2}\right)^2}
-  \left(\frac{\vartheta_1}{\eta}\right)^4
\overline{\left(\frac{\vartheta_1\theta_4}{\eta^2}\right)^2}
\right\}  \comma \\
e^{-\frac{i\pi}{2}wm} 
\left\{ 
\left(\frac{\theta_4}{\eta}\right)^4
\overline{\left(\frac{\theta_4\theta_2}{\eta^2}\right)^2}
- \left(\frac{\theta_2}{\eta}\right)^4
\overline{\left(\frac{\theta_2\theta_4}{\eta^2}\right)^2}
- \left(\frac{\theta_3}{\eta}\right)^4
\overline{\left(\frac{\theta_3\vartheta_1}{\eta^2}\right)^2}
+ \left(\frac{\vartheta_1}{\eta}\right)^4
\overline{\left(\frac{\vartheta_1\theta_3}{\eta^2}\right)^2}
\right\} \comma \\
\left|\frac{\theta_3}{\eta}\right|^8 + \left|\frac{\theta_4}{\eta}\right|^8
+\left|\frac{\theta_2}{\eta} \right|^8
+\left|\frac{\theta_1}{\eta} \right|^8
\comma 
\end{array}
\right.   \nn
\end{align}
\begin{align}
\label{fab}
& f_{(w,m)}(\tau) \\
&  :=  
\frac{1}{2} \cdot
\left\{
\begin{array}{ll}
e^{\frac{\pi i}{2}w}
\left\{
\left(\frac{\theta_3}{\eta}\right)^2\left(\frac{\theta_4}{\eta}\right)^2
- \left(\frac{\theta_4}{\eta}\right)^2\left(\frac{\theta_3}{\eta}\right)^2
- \left(\frac{\theta_2}{\eta}\right)^2\left(\frac{\vartheta_1}{\eta}\right)^2
- \left(\frac{\vartheta_1}{\eta}\right)^2\left(\frac{\theta_2}{\eta}\right)^2
\right\}  \comma \\
e^{-\frac{\pi i}{2}m}
\left\{
\left(\frac{\theta_3}{\eta}\right)^2\left(\frac{\theta_2}{\eta}\right)^2 
- \left(\frac{\theta_2}{\eta}\right)^2\left(\frac{\theta_3}{\eta}\right)^2
+ \left(\frac{\theta_4}{\eta}\right)^2\left(\frac{\vartheta_1}{\eta}\right)^2 
+ \left(\frac{\vartheta_1}{\eta}\right)^2\left(\frac{\theta_4}{\eta}\right)^2
\right\} \comma \\
  e^{-\frac{\pi i }{2}wm}
\left\{ 
\left(\frac{\theta_4}{\eta}\right)^2\left(\frac{\theta_2}{\eta}\right)^2
- \left(\frac{\theta_2}{\eta}\right)^2\left(\frac{\theta_4}{\eta}\right)^2
+\left(\frac{\theta_3}{\eta}\right)^2\left(\frac{\vartheta_1}{\eta}\right)^2
+ \left(\frac{\vartheta_1}{\eta}\right)^2\left(\frac{\theta_3}{\eta}\right)^2
\right\}  \comma \\
\left(\frac{\theta_3}{\eta}\right)^4
- \left(\frac{\theta_4}{\eta}\right)^4
-\left(\frac{\theta_2}{\eta}\right)^4
-\left(\frac{\theta_1}{\eta}\right)^4 \period
\end{array}
\right. \nn
\end{align}
The right-hand  sides of  \eqn{FabT4} and \eqn{fab} 
are listed in the order of the cases
$(w \in 2\bbZ, m \in 2 \bbZ +1)$, $(w \in 2\bbZ+1, m \in 2 \bbZ )$,
$(w \in 2\bbZ+1, m \in 2 \bbZ +1)$ and 
$(w \in 2\bbZ, m \in 2 \bbZ )$ from the top to the bottom.
$\theta_\alpha(\tau) $  $(\alpha=1,..., 4)$
and $\eta(\tau)$ are the theta functions
and the Dedekind $\eta$ function, respectively. 
We have introduced $\vartheta_1(\tau) := (-i)\theta_1(\tau)$,
and omitted the argument $\tau$ on the right-hand sides.
For the theta functions, we follow the conventions
adopted
in \cite{Satoh:2015nlc}. 

In the above, 
we have explicitly written the vanishing 
terms involving $\theta_1 \equiv 0$,
to keep track of the actual contribution from each state.
We have also kept the order of the theta functions in the products
for later use to identify the corresponding vertex operators.
For instance, 
if $r < r'$, 
$\theta_{\alpha}$ coming from the complex fermion
$\psi_R^{2r} + i \psi_R^{2r+1}$  
 is on the left side of
$\theta_{\alpha'} $ from  $\psi_R^{2r'} + i \psi_R^{2r'+1}$.
The order of $\theta_\alpha$'s in $F^{T^4}_{(w,m)}$ might not be
obvious, but can be understood from the fermionic formulation of the
$T^4_{\rm fiber}$ part.
Under the modular transformations,
\be
  S: \ \tau \ \mapsto \  
  - \frac{1}{\tau} \comma \qquad
  T: \ \tau \ \mapsto \   
  \tau + 1 \comma 
\ee
these partition functions transform covariantly as
\begin{alignat}{2}
\label{modulartransZFf}
Z_{(w,m)}^{\rm base}(\tau,\bar{\tau})|_S 
&= Z_{(m,-w)}^{\rm base}(\tau,\bar{\tau}) \comma
& \qquad
Z_{(w,m)}^{\rm base}(\tau,\bar{\tau})|_T 
&= Z_{(w,m)}^{\rm base}(\tau,\bar{\tau}) \comma \nn \\
  F^{T^4}_{(w,m)}(\tau,\bar{\tau})|_S &= F^{T^4}_{(m,-w)}(\tau,\bar{\tau}) \comma
 & F^{T^4}_{(w,m)}(\tau,\bar{\tau})|_T 
 &= F^{T^4}_{(w,w+m)}(\tau,\bar{\tau}) \comma  \\
  f_{(w,m)}(\tau)|_S &= f_{(m,-w)}(\tau) \comma
&
f_{(w,m)}(\tau)|_T 
&= 
- e^{- \frac{\pi i}{3}}
f_{(w,w+m)}(\tau) \comma \nn
\end{alignat}
where $f_{(w,m)}(\tau) \vert_T := f_{(w,m)}(\tau+1)$, and so on.


\section{Twist fields for a chiral boson}
\label{app:ATtw}

The chiral reflection $X(z) \longmapsto -X(z) $ of a chiral boson $X$
is implemented by the twist field $\Sigma(z)$ 
and its dual $\bar\Sigma(z)$ 
of the type in the Ashkin-Teller model, which have dimension $1/16$, 
and create the branch cut on the world-sheet.
These twist fields have been discussed e.g. in
 \cite{Bershadsky:1986mt,Dixon:1986qv,Zamolodchikov:1987ae,
 Ginsparg:1988ui,Hashimoto:1996he,Frohlich:1999ss,Mukhopadhyay:2001ey,
 Mattiello:2018kue}. 
 The correlation functions involving these fields 
 are, for example, 
 given by
\begin{align}
\label{twistedcorr1}
& \Big\bra \del X (z_1)  \del X(z_2) \bar\Sigma(z_3)  \Sigma(z_4) \Big\ket
 =-\frac{\alpha'}{2} \cdot\frac{1}{2}
\frac{\sqrt{\xi} + \sqrt{\xi^{-1} }}{z_{34}^{1 \over 8} z_{12}^2} 
\comma  \\
\label{twistedcorr2}
 & \Big\bra e^{ik X}(z_1) \bar\Sigma(z_2)  \Sigma(z_3) \Big\ket
 = \frac{e^{ikx_0}}{
 z_{12}^{{\alpha' \over 4} k^2} 
 z_{13}^{{\alpha' \over 4} k^2}
 z_{23}^{\frac{1}{8}-{\alpha' \over 4} k^2} } \comma \\
 \label{twistedcorr3}
& \Big\bra e^{ik_1 X}(z_1)  e^{ik_2 X}(z_2)\bar\Sigma(z_3)  
\Sigma(z_4) \Big\ket
 = \frac{e^{ix_0(k_1+k_2)}}{
 (z_{13}z_{14})^{{\alpha' \over 4} k_1^2} 
 (z_{23}z_{24})^{{\alpha' \over 4} k_2^2}
 z_{34}^{\frac{1}{8}-{\alpha' \over 4} k_1^2-{\alpha' \over 4} k_2^2} } 
  \left(\frac{1-\sqrt\xi}{1+\sqrt{\xi}}\right)^{\frac{\alpha'}{2}k_1k_2} 
  \comma 
\end{align}
where $\xi =z_{13}z_{24}/z_{23}z_{14} $, and 
$x_0$ is the zero-mode of $X$ (see e.g. \cite{Mattiello:2018kue}).
For $(z_1, z_2, z_3, z_4) = (z,w, \infty, 0)$, 
the correlator in
\eqn{twistedcorr1}
gives the two-point function in the twisted sector, 
\be 
\label{delX2pt}
\bra \Sigma \vert \del X(z) \del X(w) \vert \Sigma \ket 
  = -\frac{\alpha'}{2} \cdot\frac{1}{2}
  \frac{ \sqrt{\frac{z}{w}} + \sqrt{\frac{w}{z}}}{(z-w)^2} 
  \comma
\ee
where 
$\bra \Sigma \vert := \lim_{z\to\infty} z^{1/8} \langle 0 \vert \bar\Sigma(z)$. 
For $z \to w$, 
this takes the same form as that 
in the untwisted sector
$\big\langle \del X(z)  \del X(w) \big\rangle =  - \frac{\alpha'}{2} (z-w)^{-2}$.

The correlators involving $e^{ikX}$'s are determined 
by the  Knizhnik-Zamolodchikov(KZ)-like equation 
based  on 
the two-point function of $\del X$ for the twisted boundary condition
\eqn{delX2pt}, 
and the Sugawara-form of the energy momentum tensor,
up to the zero-mode factor 
$e^{ix_0( \cdots )}$\cite{Frohlich:1999ss}. 
The derivation relies only on general properties of the twist fields 
and not on their details.  The zero-mode factor
can be fixed by a path-integral argument as in 
\cite{Mukhopadhyay:2001ey}. 

The factors such as $z_{12}^{\alpha' k^2/4}$
in \eqn{twistedcorr2} and \eqn{twistedcorr3}
can be understood as coming from the self-contraction of $X$ in $e^{ikX}$
by adopting the same regularization as for the untwisted boundary condition
\cite{Mukhopadhyay:2001ey}. 
These factors ensure the correct scaling dimension
of the correlators.
With these remarks in mind, 
\eqn{twistedcorr3} with $(z_1, z_2, z_3, z_4) = (z,w, \infty, 0)$, 
as well as \eqn{delX2pt}, is
also understood as being obtained 
from the two-point function 
of $X$ in the twisted sector
(twisted propagator),
\be
\label{twprop}
\bra \Sigma \vert  X(z)  X(w) \vert \Sigma \ket 
  = -\frac{\alpha'}{2}  \log
  \left(\frac{\sqrt{z}-\sqrt{w}}{\sqrt{z}+\sqrt{w}}\right)
  \period
\ee

In the symmetric twist where both 
left- and right moving
bosons 
are reflected as 
$X(z) + \bar{X}(\zbar) \to - [X(z) + \bar{X}(\zbar)]$, the zero-mode
 $x_0+\bar{x}_0$ corresponds to the location of the fixed points, 
which specifies a sector among the twisted sectors. 
Each fixed point has the corresponding twist operator 
\cite{Ginsparg:1988ui}. 

We note
that the momentum conservation does not need 
to hold in \eqn{twistedcorr2}, \eqn{twistedcorr3}. 
Since the twisted sectors should be associated with the fixed points,
this may be understood also as a consequence of 
the break down of the translational invariance.
The fixed points may supply the momenta as in the case of D-branes.

\section{Formulas of integrals}
\label{app:integal}

The amplitudes in section \ref{sec:pR04pt} are obtained by the
following formula
\cite{Polchinski:1998rq,Blumenhagen:2013fgp,Kawai:1985xq}, 
\begin{align}
\label{zint}
I(\alpha,\beta;n,m) & :=
\int d^2 z \, |z|^\alpha |1-z|^\beta z^n (1-z)^m \nn \\
& = 
2\pi
\frac{\Gamma\big(1+n+\frac{\alpha}{2}\big)\Gamma\big(1+m+\frac{\beta}{2}\big)
\Gamma\big(-1-\frac{\alpha+\beta}{2}\big)}{
\Gamma\big(-\frac{\alpha}{2}\big)\Gamma\big(-\frac{\beta}{2}\big)
\Gamma\big(2+n+m+\frac{\alpha+\beta}{2}\big)} \comma
\end{align}
where $\Gamma(x)$ is the Gamma function.
This is rewritten in other forms by using 
$\Gamma(z) \Gamma(1-z) = \pi/\sin \pi z$.

In section \ref{sec:4pt2}, we need to evaluate the integral,
\be
\label{I'def}
 J(\alpha,\beta; \bar\alpha,\bar\beta;\bar\gamma)
 := \int d^2z \, z^\alpha(1-z)^\beta 
 \zbar^{\bar\alpha}(1-\zbar)^{\bar\beta} 
 \left(\frac{1-\sqrt{\zbar}}{1+\sqrt{\zbar}}\right)^{\bar\gamma}
 \period
\ee
The integral is first defined in the parameter region, in which 
the integral is convergent, and then continued to 
other regions as in \eqn{zint}.
Because of the branch cuts, the phases/branches need
to be defined appropriately. Below, we follow the standard 
procedure (see e.g. \cite{Kawai:1985xq,Dotsenko1988}).

We set $z= u_1 + i u_2$ $(u_1, u_2 \in \bbR)$. 
We also take $ |\arg z \, | < \pi$, and then
$\sqrt{z} = -1$ is out of the integration region.
The branch points on the $u_2$-plane are at
$u_2 = \pm i u_1, \pm i (u_1 -1)$. 
We deform the contour of $u_2$ along the real axis 
to the one along the imaginary axis by 
$
  u_2 \to i e^{-2i\ep} u_2 \approx i(1-2i\ep) u_2 
$
with small $ \ep > 0$.
The original $z$ and $\zbar$ become
$
 z  \to u_- + i \ep (u_+ - u_-) \comma \ 
 \zbar  \to u_+ - i \ep (u_+ - u_-) \comma
$
where  $u_\pm := u_1 \pm u_2$ with $u_1, u_2 \in \bbR$.
The integration measure is 
$
 \int d^2z = 2 i \int du_1 du_2 = i \int du_+ du_- \period
$

\begin{figure}[t]
   \centering
  \includegraphics[width=56mm]{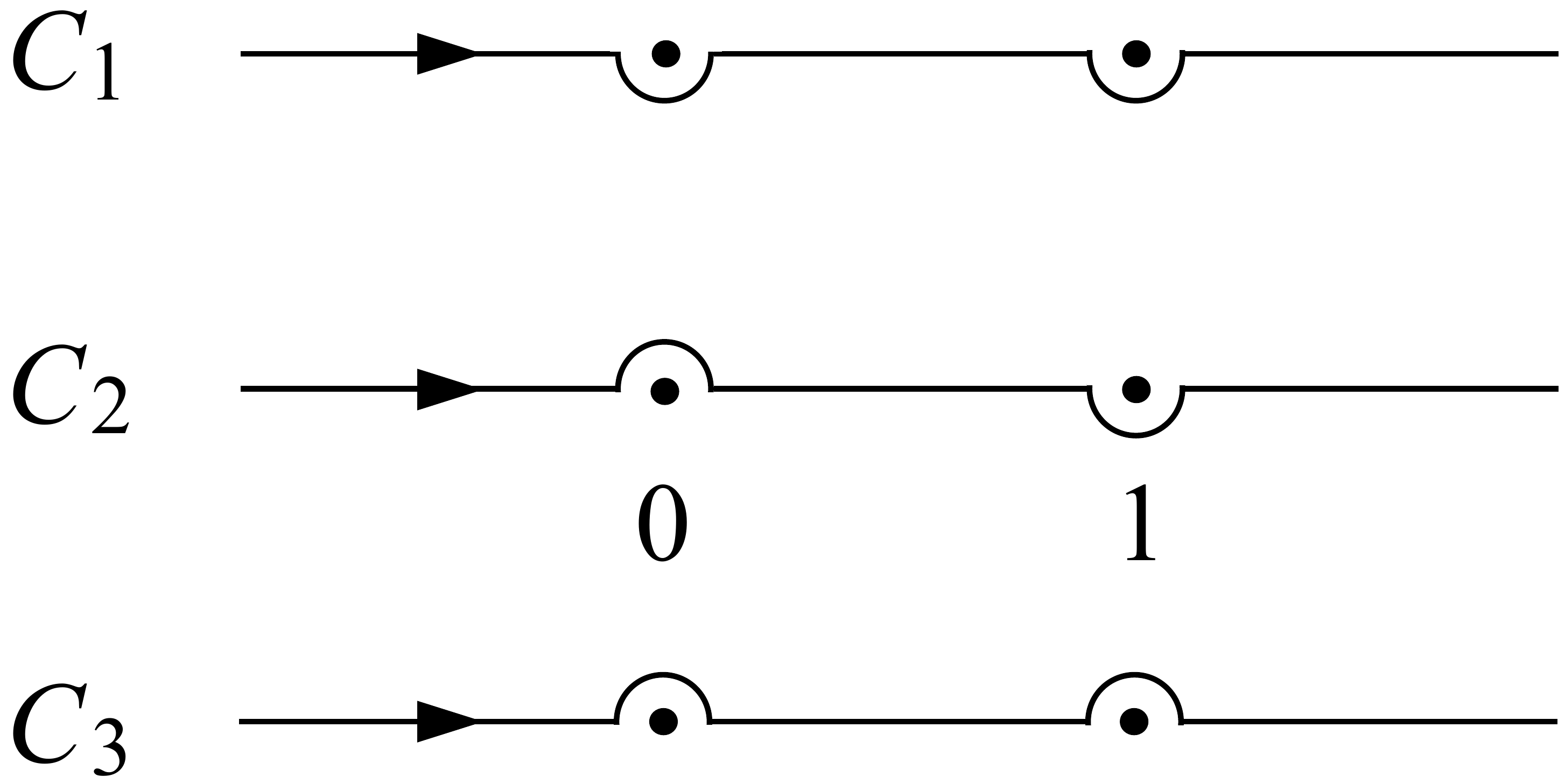} 
    \caption{Contours of $u_-$-integral. $C_1$, $C_2$, $C_3$ are
    for $u_+ < 0$,  $0< u_+ < 1$ and $ 1< u_+ $,  respectively.  }
    \label{fig:contour}
\end{figure}

The small imaginary part in $u_-$ specifies the way that the contour 
avoids the branch points. Explicitly, it is given as in Figure 
\ref{fig:contour} according to
the value of $u_+$, namely,  $u_+ < 0$, $0 < u_+ < 1$ or $1 < u_+ \,$.  
As long as the integral is 
convergent, the contributions from $u_+ < 0$ and $1 < u_+$ vanish,
which is confirmed by closing the contour in the lower/upper half plane.
Thus, we are left with
\be
  J \sim  i \int_0^1 du_+ \, 
  u_+^{\bar\alpha}
  (1-u_+)^{\bar\beta} 
 \left(\frac{1-\sqrt{u_+}}{1+\sqrt{u_+}}\right)^{\bar\gamma}
 \times
 \int_{C_2} du_- \, u_-^\alpha(1-u_-)^\beta \comma
\ee
where $C_2$ is the second contour in Figure \ref{fig:contour}.
We define the phase of the integrand for $u_- < 0$ by 
$1$ for $u_+ < 0$, $e^{ \pi i\bar\alpha}$ for $0 < u_+ < 1$
and $e^{\pi i (\bar\alpha+\bar\beta)}$ for $ 1 < u_+$.
$1, \, e^{\pi i\bar\alpha}, \, e^{\pi i(\bar\alpha+\bar\beta)}$
are understood as 
the relative phases coming from the integrand for $u_+$.
This choice/definition corresponds to choosing the phase
so that it becomes 1 when $\alpha = \bar\alpha$, 
$\beta = \bar\beta$ and $u_\pm$ are in the same 
interval of $(-\infty, 0)$, $(0,1)$ or $(1, \infty)$.
We also define $\sqrt{u_+}$
without a phase/sign for $u_+ >0$.

By deforming the contour $C_2$ for $u_-$ to be wrapped around 
$u_- = 1$, we arrive at
\be
J = -2e^{\pi i(\bar\alpha-\alpha)} \sin(\pi\beta) \times J_L \times J_R 
\comma
\ee
where
\begin{align}
\label{JLR}
J_L &=  \int_1^\infty du_- \, |u_-|^\alpha |u_- -1|^\beta
 = B(1+\beta, -\alpha-\beta-1) 
 \comma \nn \\
 J_R &= \int_0^1 du_+ \, |u_+|^{\bar\alpha}
  |1-u_+|^{\bar\beta} 
 \left(\frac{1-\sqrt{|u_+|}}{1+\sqrt{|u_+|}}\right)^{\bar\gamma}
 \comma
\end{align}
and $B(a,b)$ is  the beta function.
$J_R$ is then evaluated by making a change of variables $\sqrt{u_+} = v$, 
\begin{align}
\label{I+result}
 J_R(\bar\alpha, \bar\beta, \bar\gamma)  &= 2 \int_0^1 dv \, v^{1+2\bar\alpha} (1-v)^{\bar\beta+\bar\gamma}
 (1+v)^{\bar\beta-\bar\gamma} \nn \\
 & = 2 B(2+2\bar\alpha, 1+\bar\beta+\bar\gamma) \, 
 F(2+2\bar\alpha, \bar\gamma -\bar\beta; 
 3+2\bar\alpha+
 \bar\beta+\bar\gamma;-1) \comma
\end{align}
where 
$F(a,b;c;z)$
 is the hypergeometric function. 
 The possible poles of $J_R$ come from the factor 
 $\Gamma(2+2\bar\alpha)\Gamma(1+\bar\beta+\bar\gamma)$
 in the 
 beta function, since $\frac{1}{\Gamma(c) } F(a,b;c;z)$ is an entire function 
of $a,b,c$ for fixed $z$ with $|z| < 1$ or  $\re z < 1/2$ \cite{Bateman}. 
They can be cancelled with the zeros
from the other part.
Combining \eqn{JLR} and \eqn{I+result}, we obtain
\be
\label{I'result}
J(\alpha,\beta; \bar\alpha,\bar\beta;\bar\gamma)
= 2\pi e^{\pi i (\bar\alpha- \alpha)}
{\Gamma(-1-\alpha-\beta) \over \Gamma(-\alpha)\Gamma(-\beta)} 
J_R(\bar\alpha,\bar\beta,\bar\gamma)
\period
\ee

When $\bar\gamma =0$,  by using the formulas,
\begin{align}
 F(a,b;a-b+1;-1) = 
 {2^{-a} \sqrt{\pi} \, \Gamma(a-b+1) \over 
 \Gamma\big( \frac{a+1}{2}\big) \Gamma\big( \frac{a}{2} -b+1\big)}
 \comma  \quad \ 
  \Gamma(2z)  = \frac{2^{2z-1}}{\sqrt{\pi}} \Gamma(z) \Gamma\Big(z+\frac{1}{2}\Big)
 \comma
\end{align}
one finds that 
\be
\label{I+reduce}
 J_R(\bar\alpha,\bar\beta,0) 
=  {\Gamma(1+\bar\alpha) \Gamma(1+\bar\beta)
 \over 
 \Gamma(2+\bar\alpha+\bar\beta)
 }
  \comma
\ee
and thus
\begin{align}
\label{I'gamma0}
  J(\alpha,\beta; \bar\alpha,\bar\beta;0)
  & =2\pi e^{\pi i (\bar\alpha- \alpha)} 
  {\Gamma(-1-\alpha-\beta)\Gamma(1+\bar\alpha) \Gamma(1+\bar\beta)
 \over 
 \Gamma(-\alpha)\Gamma(-\beta)\Gamma(2+\bar\alpha+\bar\beta)} \period
\end{align}
By setting 
\be
\label{paramreplace}
 \bar\gamma = 0 \comma \quad
 \alpha \to \frac{\alpha}{2} \comma \quad
 \beta \to \frac{\beta}{2} \comma \quad
 \bar\alpha \to \frac{\alpha}{2} + n 
 \comma \quad
 \bar\beta \to \frac{\beta}{2} + m \comma
\ee
\eqn{I'gamma0} is reduced to
\eqn{zint} up to a phase $e^{\pi i (\bar\alpha- \alpha)} $ 
due to the difference
of the conventions of the phase. 
 In the main text, we drop this phase
to conform to \eqn{zint} and the definition of the phase of $\sqrt{\zbar}$
in the computation of the correlators.

The integral with the factor from
the twisted propagator also in the left-mover,
\be
 {\cal J}
  := \int d^2z \, z^\alpha(1-z)^\beta 
 \zbar^{\bar\alpha}(1-\zbar)^{\bar\beta} 
 \left(\frac{1-\sqrt{z}}{1+\sqrt{z}}\right)^{\gamma}
 \left(\frac{1-\sqrt{\zbar}}{1+\sqrt{\zbar}}\right)^{\bar\gamma}
 \comma
\ee
is evaluated similarly.
In that case, the beta function for 
$J_L$ in \eqn{JLR} is replaced by a hypergeometric function
as
\begin{align}
\label{calJ}
{\cal J} (\alpha,\beta; \bar\alpha,\bar\beta;\gamma; \bar\gamma)
&= -2e^{\pi i(\bar\alpha-\alpha)} \sin(\pi\beta) 
\times {\cal J}_L(\alpha,\beta,\gamma) 
\times J_R(\bar\alpha,\bar\beta,\bar\gamma) 
\comma \nn \\
{\cal J}_L(\alpha,\beta,\gamma) &=J_R(-\alpha -\beta -2, \beta,\gamma)
\period 
\end{align}
Compared with \eqn{I+result} for the right-mover,  
${\cal J}_L$  has 
the argument $-\alpha -\beta -2$
instead of $\alpha$ (up to the overline $\overline{(\ \ ) }$), which
implies that the roles of the $t$- and $u$-channels are exchanged.

We also list the formulas of the
hypergeometric function used in the main text
\cite{{Bateman}},
\begin{align}
\label{F1}
F(a,b;c;z) &= (1-z)^{-a} F\Big(a,c-b;c; \frac{z}{z-1}\Big)\comma  \\
\label{F2}
F\Big(a,1-a;c;\frac{1}{2}\Big) &={ 2^{1-c} \sqrt{\pi} \, \Gamma(c) 
\over \Gamma\big( \frac{c+a}{2}\big) \Gamma\big( \frac{c-a+1}{2}\big)}
\comma \\
\label{F3}
F(0,b;c;z) &=1  \comma  \\
\label{F4}
F(a,b;a;z)&= (1-z)^{-b}
\period 
\end{align}
The forth formula is obtained by combining the first and the third.

\section{so(4) gamma matrices}
\label{app:so4gamma}

We summarize our conventions of the so(4) gamma matrices 
used in section \ref{sec:4pt2}.
We denote 
by $\vert \ep_1, \ep_2 \rangle $ with $\ep_{1}, \ep_2 = \pm $
the state corresponding to the 
bosonized spinor $e^{\frac{i}{2}(\ep_1 H^1 +\ep_2 H^2 )}$ as in
\eqn{defSR}. We also set $\vert 1 \rangle = \vert +, + \rangle$,
$\vert 2 \rangle = \vert -, - \rangle$, $\vert \dot{1} \rangle = \vert +, - \rangle$,
$\vert \dot{2} \rangle = \vert +, - \rangle$.
Then, we represent  the so(4) gamma matrices
satisfying $\{ \Gamma^a, \Gamma^b \} = 2 \delta^{ab}$ by
\be
\Gamma^1 =  \sigma_1 \otimes \sigma_1
\comma \quad
\Gamma^2=  \sigma_1 \otimes \sigma_2
\comma \quad
\Gamma^3=  \sigma_1 \otimes \sigma_3
\comma \quad
\Gamma^4=   \sigma_2 \otimes {\bf 1} \comma
\ee
where $\sigma_j$ are the Pauli matrices and ${\bf 1} $ is the
$2\times2$ identity matrix. 
$ \Gamma^{j\pm} :=  (\Gamma^{2j-1} \pm i \Gamma^{2j})/2 $ $(j=1,2)$
raise or lower $\ep_j$.
The basis of the states is ordered 
as $\vert 1 \rangle, \vert 2 \rangle, \vert \dot{1} \rangle, \vert \dot{2} \rangle$.
The chirality matrix is 
$\Gamma_5 = i^2 \Gamma^1\Gamma^2 \Gamma^3\Gamma^4 =
\sigma_3 \otimes {\bf 1} $, whereas the charge conjugation matrix
is $C:= \Gamma^3 \Gamma^1 = {\bf 1} \otimes i \sigma_2$, which
satisfies $C^2 = -1$ and $C \Gamma^a C^{-1} = -(\Gamma^a)^T$.
We also use the chiral blocks of 
$\Gamma^a$, $C$ and 
$\Gamma^{ab} := (\Gamma^a\Gamma^b- \Gamma^b\Gamma^a )/2$
as in 
\be
\Gamma^a = \begin{pmatrix} 
 0 & (\gamma^a)_\alpha^{\ \dot\beta} \\
 (\bar\gamma^a)_{\dot\alpha}^{\ \beta} & 0
\end{pmatrix} \comma \quad \
C = \begin{pmatrix}
   \calC^{\alpha\beta} & 0 \\
   0 & \bar\calC^{\dot\alpha\dot\beta}
   \end{pmatrix} \comma \quad \
\Gamma^{ab} 
= \begin{pmatrix}
   (\gamma^{ab})_\alpha^{\ \beta} & 0 \\
   0 & (\bar\gamma^{ab})_{\dot\alpha}^{\ \dot\beta}
   \end{pmatrix}
    \period
\ee
In particular, $\calC^{\alpha\beta} = (i\sigma_2)^{\alpha\beta}$.
The indices are raised and lowered by $C$ and $C^{-1}$, respectively.

\vspace{4ex}
\setlength{\parskip}{0ex}

\end{document}